\documentclass[acmsmall]{acmart}
\usepackage{cases}
\usepackage{color}
\usepackage[linesnumbered, ruled]{algorithm2e}
\usepackage{subfigure}
\newtheorem{mydef}{Definition}

\newtheorem{mypro}{Problem}

\graphicspath{{./eps/}}
\newtheorem{mythm}{Theorem}
\newtheorem{myas}{Assumption}
\newtheorem{myrem}{Remark}

\newtheorem{mycol}{Corollary}

\newcommand{\rfig}[1]{Fig.\,\ref{#1}} 
 
\newcommand{\req}[1]{\eqref{#1}} 
 
\newcommand{\rtab}[1]{Table\,\ref{#1}}

\newcommand{\rcol}[1]{Corollary\,\ref{#1}}
\newcommand{\rrem}[1]{Remark\,\ref{#1}}
\newcommand{\rsec}[1]{Section\,\ref{#1}}
\newcommand{\rdef}[1]{Definition\,\ref{#1}}
\newcommand{\ras}[1]{Assumption\,\ref{#1}}
\newcommand{\ralg}[1]{Algorithm\,\ref{#1}}
\newcommand{\rpro}[1]{Problem\,\ref{#1}}
\newcommand{\rline}[1]{line\,\ref{#1}}

\newcommand{\rthm}[1]{Theorem\,\ref{#1}}

\newcommand{\qedwhite}{\hfill \ensuremath{\Box}}
\AtBeginDocument{%
  \providecommand\BibTeX{{%
    \normalfont B\kern-0.5em{\scshape i\kern-0.25em b}\kern-0.8em\TeX}}}

\usepackage[whole]{bxcjkjatype}
\usepackage{color}
\definecolor{forestgreen}{rgb}{0.33,0.61,0.34}
\definecolor{deepmagenta}{rgb}{0.8, 0.0, 0.8}
\definecolor{harvardcrimson}{rgb}{0.79, 0.0, 0.09}

\usepackage[normalem]{ulem}

\setcopyright{acmcopyright}
\copyrightyear{2021}
\acmYear{2021}
\acmDOI{10.1145/3470453}

\begin{document}

\title{Collaborative rover-copter path planning and exploration with temporal logic specifications based on Bayesian update under uncertain environments}


\author{Kazumune Hashimoto}
\affiliation{%
  \institution{Graduate School of Engineering, Osaka University}
  \city{Suita}
  \country{Japan}
}

\author{Natsuko Tsumagari}
\affiliation{%
  \institution{Graduate School of Engineering and Science, Osaka University}
  \city{Toyonaka}
  \country{Japan}
}

\author{Toshimitsu Ushio}
\affiliation{%
  \institution{Graduate School of Engineering and Science, Osaka University}
  \city{Toyonaka}
  \country{Japan}
}

\begin{abstract}
This paper investigates a collaborative rover-copter path planning and exploration with temporal logic specifications under uncertain environments. The objective of the rover is to complete a mission expressed by a syntactically co-safe linear temporal logic (scLTL) formula, while the objective of the copter is to actively explore the environment and reduce its uncertainties, aiming at assisting the rover and enhancing the efficiency of the mission completion. To formalize our approach, we first capture the environmental uncertainties by environmental beliefs of the atomic propositions, under an assumption that it is unknown which properties (or, atomic propositions) are satisfied in each area of the environment. The environmental beliefs of the atomic propositions are updated according to the Bayes rule based on the Bernoulli-type sensor measurements provided by both the rover and the copter. Then, the optimal policy for the rover is synthesized by maximizing a belief of the satisfaction of the scLTL formula through an implementation of an automata-based model checking. An exploration policy for the copter is then synthesized by employing the notion of an entropy that is evaluated based on the environmental beliefs of the atomic propositions, and a path that the rover intends to follow according to the optimal policy. As such, the copter can actively explore regions whose uncertainties are high and that are relevant to the mission completion. Finally, some numerical examples illustrate the effectiveness of the proposed approach. 
\end{abstract}




\maketitle

\section{Introduction}\label{introductionsec}
Autonomous systems play an important role to accomplish complex, high level scientific missions autonomously under uncertain environments. To increase the efficiency of completing the mission, integrating a collaboration of multiple, heterogeneous robots has attracted much attention in recent years, see, e.g., \cite{lewis2014}. In this paper, we are particularly interested in the situation, where completing the mission will be achieved by the collaboration of an unmanned ground vehicle (UGV), which is called a \textit{rover}, and an unmanned aerial vehicle (UAV), which is called a \textit{(heli)copter}. The utilization of the rover-copter collaboration is motivated by the fact that the rover has the role to complete the mission (e.g., search for a target object, etc.), while the copter has the role to \textit{assist} the rover so as to enhance the efficiency of completing the mission. Specifically, the copter aims at actively exploring the environment and reducing its uncertainties by revealing which properties (obstacles, free space, \textit{etc.}) are satisfied in each area of the environment. 
For example, the copter checks if no obstacles are present along with the path that the rover intends to follow in the environment. 
By doing so, the rover will be able to complete the mission while guaranteeing safety. Employing the rover-copter collaboration is also motivated by the fact that the National Aeronautics and Space Administration (NASA) is launching Mars 2020 mission \cite{balaram2018}. In particular, to investigate Martian geology and habitability, NASA has decided to send copters to Mars in order to help the rover discover target samples in an efficient way \cite{landau2015}. Motivated by this fact, several motion planning techniques employing the rover-copter collaboration for the Mars exploration have been investigated in recent years, see, e.g., \cite{nilsson2018,bharadwaj2018,sasaki2020}.

As briefly mentioned above, planning under environmental uncertainties has two distinct major problems; the first one is how to synthesize a control policy such that a complex, high level mission specification can be satisfied in an automatic way, and the second one is how to explore the environment so as to effectively reduce its uncertainties. In this paper, we propose a novel algorithm to solve these two problems by making use of the rover-copter collaboration. First, we tackle the former problem by employing temporal logic synthesis techniques \cite{temporalreview1,temporalreview2,belta2017formal}. \textcolor{black}{More specifically, we express a mission specification by a syntactically co-safe linear temporal logic (scLTL). In contrast to the simple reach-avoid task, the scLTL formula has the ability to describe various complex specifications that involve logic and {temporal} constraints \cite{belta2017formal}. 
Moreover, the optimal policy that fulfills the scLTL specification can be synthesized using a value iteration algorithm, which is in general more computationally efficient than synthesizing controllers with the LTL (requires to solve the Rabin game) or the STL (requires to solve the (mixed) integer programming). 
Additionally, the utilization of the scLTL is sometimes natural in practice, since the path planning problem often deals with the mission that terminates in a \textit{finite} time rather than the infinite time like LTL.} 
To formalize our approach, we first capture the uncertain environment by assuming that, it is unknown which properties or atomic propositions are satisfied in each area of the environment. Specifically, we define \textit{environmental beliefs} of the atomic propositions, which are described by posterior probabilities, that evaluate their uncertainties based on the sensor measurements in each area of the environment. 
As will be detailed later, these beliefs are updated according to the Bayes rule based on sensor measurements provided by both the copter and the rover. The optimal policy for the rover is then synthesized by maximizing a belief of the satisfaction of the scLTL formula for the controlled trajectory of the rover. 
In particular, based on an automata-based model checking (see, e.g., \cite{baier}), we combine a motion model of the rover described by a Markov decision process (MDP) and a finite state automaton (FSA) that accepts all good prefix satisfying the scLTL formula. This combined model, which is called a product belief MDP, has a transition function induced by the current environmental beliefs of the atomic propositions. The problem for finding the optimal policy is then reduced to a finite-time reachability problem in the product belief MDP, which can be solved via a value iteration algorithm. 

The latter problem (i.e., how to explore the environment so as to effectively reduce its uncertainties) will be mainly solved by the \textit{copter}, since it is able to move more quickly and freely than the rover and is thus suited for the exploration. Roughly speaking, the objective of the copter is to actively explore the environment and reduce its uncertainties by updating the environmental beliefs of the atomic propositions. We first describe the observations by employing a \textit{Bernoulli}-type sensor model, see, e.g., \cite{bertuccelli2005,wang2009,hussein2007,imai2013}. The Bernoulli sensor abstracts the complexity of image processing into binary observations. Despite the simplicity, the Bernoulli sensor model is commonly used in the UAV community, due to the following reasons; (i) it is able to capture the erroneous observations; (ii) it is able to capture the limited sensor range; (iii) in contrast to the other sophisticated sensor models such as those that involve the probability density functions, the Bayesian update can be simply computed without any integrals or approximations. 
In particular, the third feature is well-suited for the {copter's} exploration, as the computational power of the CPU and the battery capacity are often limited, and it is desirable to make the belief updates as ``computationally light'' as possible. The exploration algorithm is given by employing the notion of \textit{entropy} that is evaluated based on the environmental beliefs of the atomic propositions, and a path that the rover intends to follow according to the current optimal policy. As such, the copter can put an emphasis on actively exploring regions whose uncertainties are high and that are relevant to the mission completion. 

\subsection*{Related works and contributions of this paper}
Based on the above, the approach presented in this paper is related to the previous works of literature in terms of the following aspects: 
\begin{enumerate}
\item Motion planning/exploration employing the rover-copter collaboration; 
\item Temporal logic planning under environmental uncertainties;
\end{enumerate}
In what follows, we discuss how our approach differs from the previous works and highlight our main contributions. 

Several motion planning/exploration techniques employing the rover-copter collaboration have been provided, see, e.g., \cite{nilsson2018,bharadwaj2018,sasaki2020}. In particular, our approach is closely related to \cite{nilsson2018}, in the sense that the overall synthesis problem is decomposed into two sub-problems, i.e., the problem of synthesizing a copter's exploration policy in the uncertain environment, and the problem of synthesizing the optimal policy for the rover so as to satisfy the scLTL formula. Our approach builds upon this previous work in terms of both synthesis for the rover's optimal policy and the copter's exploration policy in the following ways. Rather than capturing the environmental uncertainties by the belief MDPs (see Section II.A in \cite{nilsson2018}), in which the environmental belief states are given in a \textit{discrete} space, this paper captures the environmental uncertainties by assigning beliefs in a \textit{continuous} space (i.e., the beliefs can take continuous values in the interval $[0, 1]$). This allows us to apply the Bayes rule to update the beliefs according to the sensor measurements provided by both the rover and the copter. 
Moreover, while the previous work solves a value iteration over the state-space in a product MDP that involves the set of environmental belief states, our approach defines a product MDP that does \textit{not} involve such states, i.e., the set of states of product MDP combines only the set of states of the MDP motion model of the rover and the set of states of the FSA that accepts all trajectories satisfying the scLTL formula (not including the environmental belief states). 
Hence, we can alleviate the time complexity of synthesizing the optimal policies for the rover in comparison with the previous work. 
\textcolor{black}{In addition, we provide a theoretical, convergence analysis of the proposed algorithm, in which the environmental beliefs of the atomic propositions are shown to converge to the appropriate values.}

Besides the above, many temporal logic planning schemes under environmental uncertainties have been proposed. 
Most of the previous works assume that the environment has unknown properties (atomic propositions) \cite{ayala2013,fu2016,maly2013,meng2013a,meng2015a,meng2018,livingston2012,wongpiromsarn2012}, or that the motion model of the robot includes uncertainties \cite{lahijanian2010,yoo2016,sadigh16,vasile2016,leahy2019,wolff2012,ulusoy2014,wongpiromsarn2012,kazumune2020}, or that the motion model of the robot is completely unknown \cite{sadigh2014,jing2015,li2018,hasanbeig2019}. 
Since this paper assumes that it is unknown which atomic propositions are satisfied in the environment, our approach is particularly related to the first category, i.e.,  \cite{ayala2013,fu2016,maly2013,meng2013a,meng2015a,meng2018,livingston2012,wongpiromsarn2012}. 
For example, \cite{ayala2013} proposed to combine an automata-based model checking and run-time verification for synthesizing a temporal logic motion planning under an incomplete knowledge about the workspace. 
\cite{fu2016} proposed a temporal logic synthesis under probabilistic semantic maps obtained by simultaneous localization and mapping (SLAM). Moreover, \cite{meng2013a} proposed a planning revision scheme under incomplete knowledge about the workspace.
Our approach is essentially different from the above previous works, in the sense that we incorporate a \textit{sensor failure} about observations on the atomic propositions. In particular, as previously mentioned, we employ a Bernoulli-type sensor model to describe erroneous observations, and update the beliefs based on the Bayes rule. Besides, the synthesis approach (e.g., construction of the product MDP) is also different from the above previous works, since we make use of the beliefs to synthesize control policies, see \rsec{missionexecutesec}. Moreover, our approach is different from the above previous works, in the sense that we incorporate an explicit algorithm for \textit{exploration}, so as to reduce environmental uncertainties. 
{Other than the above previous works, a few approaches that take into account sensor failures/noise have been provided, see, e.g., \cite{johnson2013,johnson2015,nuzzo,TIGER2020325}. 
For example, in \cite{johnson2015}, the authors proposed a probabilistic model checking for a reactive synthesis under sensor failures and actuator failures. \textcolor{black}{Moreover, \cite{nuzzo} introduced the concept of stochastic signal temporal logic (StSTL), and provided both verification and synthesis techniques using assume/guarantee contracts. In contrast to the above previous works, we here propose a \textit{Bayesian} approach, in which the beliefs that are assigned in the environment are introduced, and these are updated based on observations provided by the Bernoulli sensors.}} 

In summary, the main novelties of this paper with respect to the related works are as follows: using the copter-rover collaboration, we develop a new approach to synthesizing an optimal policy for the rover so as to satisfy an scLTL formula, and an exploration policy for the copter so as to update the environmental beliefs of the atomic propositions and reduce the environmental uncertainties. 
In particular: 
\begin{enumerate}
\item Using the Bernoulli sensor model and the Bayesian update, we propose a novel exploration algorithm for the copter so as to update the environmental beliefs and effectively reduce the environmental uncertainties (for details, see \rsec{explorationsec}); 
\item We propose a novel framework to synthesize the optimal policy for the rover. In particular, we solve a value iteration over a product MDP, whose state-space does not involve the set of the states of the environmental beliefs. This leads to the reduction of time complexity of the value iteration in comparisons with the previous work (for details, see \rsec{missionexecutesec}); 
\item \textcolor{black}{We provide a theoretical, convergence analysis of the proposed algorithm, where it is shown that the environmental beliefs of the atomic propositions converge to the appropriate values (for details, see \rsec{convergencesec}).}
\end{enumerate}

The remainder of this paper is organized as follows. In \rsec{preliminariessec}, we provide some preliminaries of the Markov decision process and syntacticallly co-safe LTL formula.
In \rsec{problemformulationsec}, we formulate a problem that we seek to solve in this paper. In \rsec{mainapproach}, we describe the main algorithm that aims to synthesize both an exploration policy to update the environmental beliefs of the atomic propositions and the optimal policy to satisfy an scLTL formula. 
In \rsec{convergencesec}, we analyze the convergence property of the proposed algorithm.
In \rsec{simulationsec}, we illustrate the effectiveness of the proposed approach through a simulation example. We finally conclude in \rsec{conclusionsec}. \\ 

\noindent
\textit{Notation.} Let $\mathbb{N}$, $\mathbb{N}_{\geq 0}$, $\mathbb{N}_{>0}$, $\mathbb{N}_{a: b}$ be the set of integers, non-negative integers, positive integers, and the set of integers in the interval $[a, b]$, respectively. 
Let $\mathbb{R}$, $\mathbb{R}_{\geq 0}$, $\mathbb{R}_{>0}$, $\mathbb{R}_{a:b}$ be the set of reals, non-negative reals, positive reals, and the set of reals in the interval $[a, b]$, respectively. For a given vector $x\in\mathbb{R}^n$, denote by $x^{(i)}$ the $i$-th element of $x$. Given a finite set $X$, let $\mathcal{D}(X)$ denote the set of all probability distributions on $X$, i.e., the set of all functions $p : X \rightarrow [0, 1]$ such that $\sum_{x \in X} p(x) = 1$. 

\section{Preliminaries}\label{preliminariessec}
\subsection{Markov Decision Process}
A Markov Decision Process (MDP) is defined as a tuple $\mathcal{M} = (X, x_{0}, U, p)$, where $X$ is the finite set of states, $x_0 \in X$ is the initial state, $U$ is the finite set of control inputs, and $p : X \times U \rightarrow \mathcal{D}(X)$ is the transition probability function that associates, for each state $x\in X$ and input $u\in U$, the corresponding probability distribution over $X$. For simplicity of presentation, we abbreviate $p(x, u) (x')$ as $p(x' | x, u)$. 
Given $\mathcal{M}= (X, x_{0}, U, p)$, a \textit{policy sequence} ${\mu}_{seq} = \mu_1 \mu_2 \ldots$ is defined as an infinite sequence of the mappings $\mu_k : X \rightarrow U$, $k\in \mathbb{N}$. Namely, each $\mu_k$, $k\in\mathbb{N}_{\geq 0}$ represents a policy as a mapping from each state in $X$ onto the corresponding control input in $U$. The policy sequence ${\mu}_{seq}$ is called \textit{stationary} if the policy is the invariant for all times, i.e., $\mu_k = \mu_{k+1}$, $\forall k \in\mathbb{N}_{\geq 0}$. Given a policy sequence ${\mu}_{seq} = \mu_1 \mu_2 \ldots$, a \textit{trajectory} induced by ${\mu}_{seq}$ is denoted by ${\bf x}_{{\mu}_{seq}} = x(0) x(1) \ldots \in X^\omega$, where $x(0) = x_0$ and $x(k+1) \sim p (\cdot | x(k), \mu_k (x(k)))$, $\forall k\in\mathbb{N}_{\geq 0}$. 

\subsection{Syntactically co-safe LTL}\label{scltlsec}
Syntactically co-safe LTL (scLTL for short) is defined by using the set of atomic propositions $AP$, Boolean operators, and some temporal operators. Atomic propositions are the Boolean variables taking either true or false. Specifically, the syntax of the scLTL formulas are constructed according to the following grammar: 
\begin{equation}\label{syntax}
\phi ::= {\rm true}\ |\ ap\ | \neg ap\ |\ \phi_1 \wedge \phi_2 \ |\ \phi_1 \vee	 \phi_2 \ |\ \bigcirc \phi \ |\ \phi_1 \mathit{U} \phi_2, 
\end{equation}
where $ap \in AP$ is the atomic proposition, $\neg$ (\textit{negation}), $\wedge$ (\textit{conjunction}), $\vee$ (\textit{disjunction}) are the Boolean connectives, and $\bigcirc$ (\textit{next}), $\mathit{U}$ (\textit{until}) are the temporal operators. The semantics of LTL formula is inductively defined over an infinite sequence of sets of atomic propositions ${\bf w} = w_0 w_1 \cdots \in (2^{AP})^\omega$. Intuitively, an atomic proposition $ap \in AP$ is satisfied iff $ap$ is true at $w_0$ (i.e., $ap \in w_0$). Moreover, $\neg ap$ is satisfied iff $ap$ is not true at $w_0$ (i.e., $ap \notin w_0$). $\phi_1 \wedge \phi_2$ is satisfied iff both $\phi_1$ and $\phi_2$ are satisfied. $\phi_1 \vee \phi_2$ is satisfied iff $\phi_1$ or $\phi_2$ are satisfied. $\bigcirc \phi$ is satisfied iff $\phi$ is satisfied for the suffix of ${\bf w}$ that begins from the next position, i.e., i.e., $w_1 w_2 \cdots$. Finally, $\phi_1 \mathit{U} \phi_2$ is satisfied iff $\phi_1$ is satisfied until $\phi_2$ is satisfied. Given ${\bf w} = w_0 w_1 \cdots \in (2^{AP})^\omega$ and an scLTL formula $\phi$, we denote by ${\bf w} \models \phi$ iff ${\bf w}$ satisfies $\phi$. It is known that every ${\bf w} = w_0 w_1 \cdots$ that satisfies the scLTL formula $\phi$ contains a finite \textit{good prefix} $w_0 w_1 \cdots w_n$ for some $n\in\mathbb{N}_{\geq 0}$, such that $w_0 w_1 \cdots w_n {\bf w}' \in (2^{AP})^\omega$ also satisfies $\phi$ for all ${\bf w}' \in (2^{AP})^\omega$.

A finite state automaton (FSA) is defined as a tuple $\mathcal{A} = (Q, \Sigma, \delta, q_0, Q_f)$, where $Q$ is a set of states, $\Sigma$ is the input alphabet, $\delta: Q \times \Sigma \rightarrow Q$ is the transition function, $q_0 \in Q$ is the initial state, and $Q_f \subseteq Q$ is the set of accepting states. Moreover, denote by $Post : Q \rightarrow 2^Q$ the successors for each state in $Q$, i.e., $Post(q) = \{q' \in Q\ |\ \exists \sigma \in \Sigma, q' \in \delta (q, \sigma)\}$. 
It is known that any scLTL formula $\phi$ can be translated into the FSA with $\Sigma = 2^{AP}$, in the sense that all good prefix for $\phi$ can be accepted by $\mathcal{A}_{\phi}$. We denote by $\mathcal{A}_{\phi}$ the FSA corresponding to the scLTL formula $\phi$. The translation from scLTL formulas to the FSA can be automatically done using several off-the-shelf tools, such as SCHECK2 \cite{latvala2003}. 

\section{Problem formulation}\label{problemformulationsec}
In this section we describe an uncertain environment, motion and sensor models of the rover and the copter, and the main problem that we seek to solve in this paper. 
\subsection{Uncertain environment}
We capture an environment as a two-dimensional map consisting of $n$ cells. For example, this map is obtained by discretizing a given bounded search area into uniform grids with $n$ cells. Let $x_i \in \mathbb{R}^2$, $i\in \{1, \ldots, n\}$ be the position or the centroid of the cell $i$, and let $X = \{x_1, \ldots, x_n\}$. Moreover, we denote by $AP$ the set of \textit{atomic propositions}, which represents a set of labels or {properties} that can be satisfied in the states. 
In addition, we denote by ${L}: X \rightarrow 2^{AP}$ the \textit{labeling function}, which represents a mapping from each state $x \in X$ onto the corresponding set of atomic propositions that are satisfied in $x$. For example, if $L(x) = \{\mathit{obstacle}\}$ with $AP = \{ \mathit{obstacle} \}$, it intuitively means that there is an obstacle in the state $x$. 
In this paper, it is assumed that the labeling function $L$ is \textit{unknown} due to the uncertainty of the environment, i.e., we do not have a complete knowledge about the properties of states in the environment. Thus, instead of $L$, we make use of the \textit{belief} or the posterior probability (given the past observations) as follows: 
\begin{align}\label{beliefatomic}
\mathcal{B} (x\models ap)\in[0,1] 
\end{align}
for all $x \in X$ and $ap\in AP$, where $x\models ap$ denotes that $ap$ is satisfied in $x$ (i.e., $ap \in L(x)$). 
For example, $\mathcal{B}(x\models \mathit{obstacle}) = 1$ with $AP = \{\mathit{obstacle}\}$ intuitively means that, it is for sure that there exists an obstacle in $x$. 
In addition, $\mathcal{B}(x\models \mathit{obstacle}) = 0.5$ intuitively means that, it is completely unknown whether there exists an obstacle in $x$. The belief that $ap$ is not satisfied in $x$ is denoted as $\mathcal{B} (x \models \neg ap)$, and, from \req{beliefatomic}, it is computed as $\mathcal{B} (x \models \neg ap) = 1-\mathcal{B}(x \models ap)$. 
In what follows, the beliefs in \req{beliefatomic} are called the \textit{environmental beliefs} of the atomic propositions. As we will see later, the environmental beliefs of the atomic propositions are updated based on the observations provided by sensors equipped with the copter and the rover. 

\subsection{Rover and copter model}
\subsubsection{Motion model} The motion of the rover is modeled by an MDP $\mathcal{M}_r = (X, x_{r_0}, U_r, p_r)$, where $X$ is the set of states (or the environment), $x_{r_0} \in X$ is the initial state of the rover, $U_r$ is the finite set of inputs, and $p_r$ is the transition probability function. 
Similarly, the motion of the copter is modeled by an MDP $\mathcal{M}_c = (X, x_{c_0}, U_c, p_c)$, where $X$ is the set of states, $x_{c_0} \in X$ is the initial state of the copter, $U_c$ is the finite set of inputs, and $p_c$ is the transition probability function. 


\subsubsection{Sensor model} \label{sensormodel}
The rover is equipped with \textit{sensors} that can provide observations on several atomic propositions in $AP$. Specifically, let $AP_r \subseteq AP$ be a set of atomic propositions or the properties that can be observed by the rover's sensors. 
For example, if $AP_r = \{ \mathit{target} \}$, the rover is equipped with a sensor that can detect a target object. 
To describe the erroneous observations, we use a \textit{Bernoulli}-type sensor model (see, e.g. \cite{bertuccelli2005,wang2009,hussein2007,imai2013}) as follows. Suppose that the rover's position is $x \in X$, and we would like the rover to know whether an atomic proposition $ap \in AP_r$ is satisfied at the position $x' \in X$. Due to the fact that the rover can provide sensor measurements only in a limited range, it is assumed that $\|x - x'\| \leq R^r _{ap}$, where $R^r _{ap} \in \mathbb{R}_{>0}$ is a given sensor range for $ap$. 
The corresponding observation is described by the binary variable, which is denoted by $Z^r _x (x' \models ap) \in \{0, 1\}$. 
For example, if $Z^r _x (x' \models \mathit{obstacle}) = 1$ with $AP_r = \{ \mathit{obstacle} \}$, the rover at the position $x$ detects an obstacle at $x'$ by the corresponding sensor. The conditional probabilities that the sensor provides the correct or the false measurement are characterized as follows: 
\begin{align*}
&{\rm Pr}\left[Z^r _x (x' \models ap) = 1 | x' \models ap \right] = \beta^r _{1, x} (x', ap), \  {\rm Pr}\left[Z^r _x (x' \models ap) = 0 | x' \models ap \right] = 1-\beta^r _{1, x} (x', ap) \\ 
&{\rm Pr}\left[Z^r _x (x'\models ap) = 0 | x' \models \neg ap \right] = \beta^r _{2, x} (x', ap),\ {\rm Pr}\left[Z^r _x (x'\models ap) = 1 | x' \models \neg ap \right] = 1- \beta^r _{2, x} (x', ap) 
\end{align*}
where $\beta^r _{1, x} (x', ap), \beta^r _{2, x} (x', ap) \in [0, 1]$ are given parameters that characterize the precision of the sensor. For example, under the fact that $ap$ is satisfied in $x'$ (i.e., $ap \in L(x')$), the probability of making the correct measurement (i.e., $Z^r _x (x' \models ap) = 1$) is $\beta^r _{1, x} (x', ap)$. On the other hand, the probability of making the \textit{false} measurement (i.e., $Z^r _x (x' \models ap) = 0$) is $1-\beta^r _{1, x} (x', ap)$. 
For simplicity, it is assumed that the probabilities of making the correct measurements are the same, i.e., $\beta^r _{1, x} (x', ap) = \beta^r _{2, x} (x', ap) = \beta^r _{x} (x', ap)$, $\forall x, x' \in X$ and $ap \in AP_r$. Regarding $\beta^r _{x}$, we assume that it is characterized by the fourth-order polynomial function of $\|x- x'\|$ as follows \cite{wang2009,hussein2007}: 
\begin{numcases}
{\beta^r _{x} (x', ap) = }\frac{M^r _{ap}}{(R^r _{ap})^4} \left(\|x- x'\|^2 - (R^r _{ap})^2\right)^2 + 0.5,\ \ {\rm if} \ \|x- x'\| \leq R^r _{ap}, \label{beta1} \\ 
0.5 \ \ \ \ \ \ \ \ \ \ \ \ \ \ \ \ \ \ \ \ \ \ \ \ \ \ \ \ \ \ \ \ \ \ \ \ \ \ \ \ \ \ \ \ \ \ \ \ \ \ \ \ \ \ \ {\rm if} \ \|x- x'\| > R ^r _{ap}, \label{beta2}
\end{numcases}
for given $M^r _{ap} \in (0, 0.5]$. \req{beta1} and \req{beta2} imply that the reliability of the sensor decreases (and eventually becomes $0.5$) as the distance between $x$ and $x'$ becomes larger. 

Similarly, let $AP_c \subseteq AP$ be a set of atomic propositions or the properties that can be observed by the copter's sensors. For simplicity, it is assumed that $AP_r \cup AP_c = AP$. Moreover, we denote by $R^c _{ap} \in \mathbb{R}_{>0}$ for each $ap \in AP_c$ a given sensor range for $ap$. Suppose that the copter's position is $x \in X$, and we would like the copter to know whether an atomic proposition $ap \in AP_c$ is satisfied at the position $x' \in X$ with $\|x'-x\| \leq R^c _{ap}$. The corresponding observation is denoted by $Z^c _x (x' \models ap) \in \{0, 1\}$. Moreover, denote by $\beta^c _{x} (x', ap) \in [0, 1]$ a given parameter that characterizes the precision of the sensor for given $M^r _{ap} \in (0, 0.5]$ as with \req{beta1} and \req{beta2}. 

Despite simplicity, the Bernoulli sensor model is commonly used in the UAV community, due to the following reasons; (i) it is able to capture the erroneous observations; (ii) it is able to capture the limited sensor range; (iii) in contrast to the other sophisticated sensor models such as those that involve the probability density functions, the Bayesian update can be simply computed without any integrals or approximations. In particular, the third feature is well-suited for the \textit{copter's} exploration, as the computational power of the CPU and the battery capacity are often limited, and it is desirable to make the belief update as computationally light as possible.

\subsection{Mission specification and problem formulation}
The mission specification that the rover should satisfy is expressed by an scLTL, denoted by $\phi$, over the set of atomic propositions $AP$. 
The satisfaction relation of the scLTL formula $\phi$ is given over the word generated by the trajectory of the rover. That is, given a policy sequence $\mu_{r,seq} = \mu_{r,0}\mu_{r,1}\mu_{r,2}\ldots$, we say that the trajectory ${\bf x}_{\mu_{r,seq}} = x(0) x(1) \ldots \in X^{\omega}$ satisfies $\phi$, which we denote by ${\bf x}_{\mu_{r,seq}} \models \phi$, iff the corresponding word satisfies $\phi$, i.e., ${\bf w} = L(x(0)) L(x(1)) \ldots \models \phi$. Since the rover aims at achieving the satisfaction of $\phi$, we would like to derive an optimal policy, such that the probability of satisfying $\phi$, i.e., ${\rm Pr} [{\bf x}_{\mu_{r,seq}} \models \phi]$, is maximized. 
However, since the labeling function $L$ is unknown, the values of ${\rm Pr} [{\bf x}_{\mu_{r,seq}} \models \phi]$ are also unknown (i.e., we do not have direct access to this probability value). Hence, we will instead compute and maximize the \textit{belief} that the trajectory of the rover satisfies $\phi$, which we denote by 
\begin{align}\label{beliefsatisfaction}
\mathcal{B} ({\bf x}_{\mu_{r,seq}} \models \phi) \in [0, 1].  
\end{align}
That is, $\mathcal{B} ({\bf x}_{\mu_{r,seq}} \models \phi)$ indicates the posterior that the trajectory of the rover satisfies $\phi$ given the past observations (sensor measurements) provided by the rover and the copter. 
As we will see later, \req{beliefsatisfaction} is computed and maximized based on the environmental beliefs of the atomic propositions in \req{beliefatomic}. Since the environmental beliefs of the atomic propositions will be updated based on the sensor measurements, the optimal policy that maximizes \req{beliefsatisfaction} will be also updated accordingly. Moreover, since we would like to reduce the environmental uncertainties as much as possible (i.e., we would like to make $\mathcal{B}(x\models ap)$ converge to $1$ or $0$ for all $x \in X, ap\in AP$), it is also necessary to explore the state space $X$ so as to collect the sensor measurements and update the environmental beliefs of the atomic propositions. 
In this paper, the \textit{copter} has the main role to explore the uncertain environment, since, as previously mentioned in \rsec{introductionsec}, it is able to move more quickly and freely than the rover. 
Therefore, we need to synthesize not only an optimal policy for the rover such that \req{beliefsatisfaction} is maximized so as to increase the possibility to satisfy $\phi$, but also an \textit{exploration policy} for the copter so as to update the environmental beliefs of the atomic propositions and effectively reduce the environmental uncertainties. 
\begin{mypro}\label{problem}
\normalfont
\noindent
{Consider} the MDP motion models of the rover $\mathcal{M}_r$ and of the copter $\mathcal{M}_c$, the Bernoulli sensor models as described in \rsec{sensormodel}, and mission specification expressed by the scLTL formula $\phi$. Then, synthesize for the copter-rover team a policy to increase the possibility to achieve the satisfaction of $\phi$. Specifically, synthesize an optimal policy for the rover such that \req{beliefsatisfaction} is maximized, and an {exploration policy} for the copter so as to update the environmental beliefs of the atomic propositions in \req{beliefatomic}. \qedwhite 
\end{mypro}

\section{Approach}\label{mainapproach}
In this section we provide a solution approach to \rpro{problem}. In \rsec{overviewsec}, we provide the overview of the approach. Then, we provide the algorithms of the exploration phase and the mission execution phase in \rsec{explorationsec} and \rsec{missionexecutesec}, respectively. 

\subsection{Overview of the approach} \label{overviewsec}
Following \cite{nilsson2018}, we consider a \textit{sequential} approach to solve \rpro{problem}. The overview of the approach is shown in \ralg{alg}. 
\begin{algorithm}[t]
{ 
$k \leftarrow 0$ (initialize the time);\ $x_{c} \leftarrow x_{c_0}$ (initialize the position of the copter);\ $x_{r} \leftarrow x_{r_0}$ (initialize the position of the rover); \\
Using prior knowledge, initialize $\mathcal{B}( x\models ap) \in (0, 1)$ for all $x \in X$ and $ap \in AP$; \\
The rover computes the optimal policy such that \req{beliefsatisfaction} is maximized, and compute the mapping $b_{\max} : X\rightarrow [0, 1]$ (for details, see \rsec{missionexecutesec}); \label{inialpolicy}\\

\smallskip 
\smallskip
{\textbf{\underline{Repeat the following two phases}}: \\  
\begin{enumerate}
\item {{\textit{{{Exploration phase (see \rsec{explorationsec}):}}}}} The copter explores the state space for a given time period $T_c$ and update the environmental beliefs of the atomic propositions $\mathcal{B}$:  
\begin{align}\label{explorationeq}
\{x_c, \mathcal{B}\} \leftarrow \mathit{Exploration}(x_c, \mathcal{B}, b_{\max}, T_c). 
\end{align}
Set $k \leftarrow k+T_c$ and the copter transmits $\mathcal{B}$ to the rover; 

\item{{\textit{{{Mission execution phase (see \rsec{missionexecutesec}):}}}}} \label{impstart} The rover computes the optimal policy such that \req{beliefsatisfaction} is maximized and executes it for a given time period $T_r$. Moreover, update the environmental beliefs of the atomic propositions $\mathcal{B}$ as well as the mapping $b_{\max}$, i.e., 
\begin{align}\label{missionexecuteeq}
\{x_r, b_{\max}, \mathcal{B}\} \leftarrow \mathit{MissionExecution} (x_r, \mathcal{B}, T_r). 
\end{align}
Set $k \leftarrow k+ T_r$ and the rover transmits $\mathcal{B}$ and $b_{\max}$ to the copter; 
\end{enumerate}
}
    \caption{Overview of the main algorithm.} \label{alg}
    }
\end{algorithm} 
With a slight abuse of notation, we denote by $\mathcal{B}$ in the algorithm the set of all environmental beliefs of the atomic propositions, i.e., $\mathcal{B}(x \models ap), x \in X, ap \in AP$. Specifically, the approach mainly consists of the two phases: the \textit{exploration} phase and the \textit{mission execution} phase. During the exploration phase, the copter explores the state-space $X$ so as to update $\mathcal{B}$ for a given time period $T_c \in \mathbb{N}_{>0}$. In \req{explorationeq} (as well as \req{missionexecuteeq}), $b_{\max} : X \rightarrow [0, 1]$ will denote a mapping from each state onto the corresponding {maximum belief} that the rover will reach within the time period $T_r$ according to the current optimal policy; for the detailed definition, see \rsec{maxbelief}. As we will see later, the exploration policy is given by making use of the mapping $b_{\max}$ and the {entropy} that will be derived from the current environmental beliefs of the atomic propositions $\mathcal{B}$. 
Once the exploration is done, the copter transmits the updated environmental beliefs of the atomic propositions $\mathcal{B}$ to the rover and moves on to the mission execution phase. During the mission execution phase, the rover computes the optimal policy such that \req{beliefsatisfaction} is maximized, and executes the policy for a given time period $T_r \in \mathbb{N}_{>0}$. Moreover, during the execution, the rover provides sensor measurements to update $\mathcal{B}$. Once the execution is done, the rover transmits the updated $\mathcal{B}$ and $b_{\max}$ to the copter and moves back to the exploration phase. 
\textcolor{black}{The sequential approach as above is motivated by the fact that, before allowing the rover to execute the optimal policy, we can let the copter in advance explore regions around the rover's (future) path. For example, the copter checks if no obstacles are present along with the path that the rover intends to follow in the future. Then, if the copter finds some obstacles in the path, the rover can re-design the path for avoiding the obstacles and try to find another way to complete the mission. Such scheme is somewhat too careful, but may be necessary to be done especially for safety critical systems, such as the exploration on Mars.} 

\textcolor{black}{As detailed below,  
the algorithms for both the exploration and the mission execution phases are significantly different from \cite{nilsson2018} in the following three aspects.
First, the environmental beliefs of the atomic propositions are updated based on the Bayes rule using the past sensor measurements. 
This allows us to provide novel copter's explorations, in which the copter actively explores the environment by evaluating both the level of uncertainty and the relevancy to the mission completion (see \rsec{explorationsec}). Second, we propose a novel framework to synthesize the optimal policy for the rover. In particular, we solve a value iteration over a product MDP, whose state-space does \textit{not} involve the set of the states for the environmental beliefs. This leads to the reduction of the time complexity of solving the value iteration algorithm (for details, see \rsec{missionexecutesec}). 
Finally, we provide a theoretical, convergence analysis of the proposed algorithm, where it is shown that the environmental beliefs of the atomic propositions converge to the appropriate values (for details, see \rsec{convergencesec}). }


\subsection{Exploration phase} \label{explorationsec}
In this subsection, we propose an algorithm of how the copter explores the environment so as to effectively update the environmental beliefs of the atomic propositions. 

\subsubsection{Bayesian belief update} \label{explorationbayesiansec}
Using the sensor model described in \rsec{sensormodel}, the copter updates the environmental beliefs of the atomic propositions based on the Bayes filter \cite{wang2009}. 
Suppose that the copter is in the position $x \in X$ and, for some $x' \in X$ and $ap \in AP_r$ with $\|x-x'\|\leq R^c _{ap}$, it gives the corresponding observation as $Z^c _x (x' \models ap) = z \in \{0, 1\}$. Then, using this observation, the belief that $ap$ is satisfied in $x'$, i.e., $\mathcal{B}(x' \models ap)$ is updated by applying the \textit{Bayes rule} as follows: 
\begin{align}\label{bayes}
\mathcal{B}(x' \models ap) \longleftarrow \cfrac{{\rm Pr}[Z^c _x (x'\models ap) = z | x' \models ap]\mathcal{B}(x' \models ap)}{{\rm Pr}[Z^c _x (x' \models ap) = z]} 
\end{align}
where ${\rm Pr}[Z^c _x (x'\models ap) = z | x' \models ap] ={\beta^c _{ x} (x', ap)}^z (1-{\beta^c _{x} (x', ap)})^{1-z}$, and ${\rm Pr}[Z^c _x (x'\models ap) = z]$ is computed as 
\begin{align}
{\rm Pr}[Z^c _x (x'\models ap) = z] &= {\rm Pr}[Z^c _x (x' \models ap) = z | x' \models ap]\mathcal{B}(x' \models ap)  \notag \\ 
&\ \ \ \ + {\rm Pr}[Z^c _x (x' \models ap) = z | x' \models \neg ap](1-\mathcal{B}(x' \models ap))\notag \\ 
&={\beta^c _{x} (x', ap)}^z (1-{\beta^c _{x} (x', ap)})^{1-z} \mathcal{B}(x' \models ap)\notag \\ 
&\ \ \ \ +{\beta^c _{x} (x', ap)}^{1-z} (1-{\beta^c _{x} (x', ap)})^z (1-\mathcal{B}(x' \models ap)).\notag 
\end{align}
As will be clearer in the overall exploration algorithm given below, if the copter is placed at $x\in X$, it obtains the sensor measurements for all its neighbors, i.e., $x' \in X$ with $\|x-x'\|\leq R_c$ and for all atomic propositions $ap \in AP_c$, and update the corresponding environmental beliefs of the atomic propositions according to \req{bayes}. 

\subsubsection{Acquisition function for exploration}\label{decisionmaking}
Let us now define an acquisition function to be evaluated for synthesizing the exploration strategy for the copter. First, we define the notion of an \textit{entropy} $H : [0, 1] \rightarrow [0, 1]$ as follows \cite{cover2006}: 
\begin{align}\label{entropy}
H(b) = - b \log b - (1-  b) \log (1- b)
\end{align}
for $b \in [0, 1]$, where $\log$ is to the base $2$. 
In essence, $H(\mathcal{B} (x \models ap))$ for $x \in X, ap \in AP$ represents the level of uncertainty about whether $ap$ is satisfied in $x$, and takes the largest value if $\mathcal{B} (x \models ap) = 0.5$ and the lowest value if $\mathcal{B} (x \models ap) = 0$ or $1$. Hence, by actively exploring the state space where the entropy is large and updating the corresponding beliefs according to \req{bayes}, it is expected that the environmental uncertainties can be effectively reduced. 

However, if the copter would explore the environment only by evaluating the above entropy, it might happen to explore the states that are completely \textit{irrelevant} to the rover's mission completion. In other words, since the copter knows the path that the rover intends to follow according to the current optimal policy, it is preferable that the copter should investigate areas around such path (before the rover executes it) so as to update the environmental beliefs of the atomic propositions. 
For example, the copter checks if no obstacles are present along with the path that the rover intends to follow in the environment, so that the rover will be able to complete the mission while avoiding any obstacles. 
In order to incorporate the rover's path for exploration, recall that the current position of the rover is $x_r$ and we have the mapping $b_{\max}:X\rightarrow [0, 1]$ (see \ralg{alg}). As previously described in \rsec{overviewsec}, $b_{\max}(x)$ for each $x\in X$ indicates the belief that the rover will reach $x \in X$ from $x_r$ within the time period $T_r$ according to the rover's current optimal policy (for the detailed definition and the calculation on $b_{\max}$, see the mission execution phase in \rsec{maxbelief}). Hence, for a large value of $b_{\max}(x)$ ($b_{\max}(x) \approx 1$), we have a high belief that the rover will reach $x$ within the time period $T_r$. Combining the entropy \req{entropy} and $b_{\max}$, let us define the \textit{acquisition function} $W : X \rightarrow \mathbb{R}_{>0}$ as follows: 
\begin{align}\label{acquisition}
W(x) = \sum_{ap \in AP_c} H(\mathcal{B} (x \models ap)) + \alpha b_{\max}(x), 
\end{align}
for all $x \in X$, where $\alpha \in \mathbb{R}_{>0}$ is the weight associated to $b_{\max}(x)$.
\subsubsection{Exploration algorithm}\label{overallexplorationsec}
We now propose an exploration algorithm. In the following, we provide two different exploration strategies so as to take both the efficiency of computation and the coverage of exploration into account. 

\begin{algorithm}[t]
{ 
\SetKwInOut{Input}{Input}
\SetKwInOut{Output}{Output}
\Input{$x$ (current copter's position); $\mathcal{B}$ (current environmental beliefs of the atomic propositions); $b_{\max}$ (mapping to indicate reachability probability of the rover); $T_c$ (time period for exploration);} 
\Output{$x$ (updated current position); $\mathcal{B}$ (updated environmental beliefs of the atomic propositions);}
\For {$\ell = 0:T_c -1$}{
Compute $H$ and $W$ according to \req{entropy} and \req{acquisition}, respectively; \label{start} \\ 
From \req{explorationpolicy}, compute the control input \textcolor{black}{$u^* _c = \mu^* _{c_1} (x)$};  \label{end}

Apply $u^* _c$ and sample the next state $x_{next} \sim p_c (\cdot |x, u^* _c)$;\\
$x \leftarrow x_{next}$; \\
\For{each $(x', ap) \in X \times AP_c$ with $\|x - x'\| \leq R^c _{ap}$\label{start1}}{
Provide the corresponding observation: $Z^c _x (x' \models ap) = z$; \\
Update $\mathcal{B}(x' \models ap)$ \textcolor{black}{according to \req{bayes}.} 
}\label{end1}
}
\Return $x$, $\mathcal{B}$; 
    \caption{$\mathit{Exploration}(x, \mathcal{B}, b_{\max}, T_c)$ (local selection-based exploration)} \label{explorationalg}
    }
\end{algorithm}

\textit{(Local selection-based policy):}  
The first exploration strategy is the \textit{local selection-based policy}, in which the copter executes a one step greedy exploration:
\begin{align}\label{explorationpolicy}
\mu^* _{c_1} (x) \in  \arg\max_{u \in U_c}\ \mathbb{E}_{x' \sim p_c (\cdot |x,u)}[W(x') | x, u]
\end{align}
for all $x \in X$. \req{explorationpolicy} implies that the copter greedy selects a control input such that the corresponding next state provides the highest acquisition. 
The overall exploration algorithm based on the local selection-based policy is summarized in \ralg{explorationalg}. As shown in the algorithm, for each step $\ell$, the copter computes the acquisition function and a control input $u^* _c \in U_c$ according to \rsec{decisionmaking} (line~6--\rline{end}). Once $u^* _c$ is obtained, the copter applies it and samples the next state $x_{next}$. Given the new current position, the copter makes the new sensor measurements for all its neighbors and the atomic propositions $AP_c$, and update the corresponding environmental beliefs of the atomic propositions (\rline{start1}--\rline{end1}). The above procedure is iterated for the copter's time period $T_c$. Note that, in order to enhance the exploration, the acquisition function as well as the policy computed in \req{explorationpolicy} are updated for each time when the copter obtains new observations. 

The local selection-based policy is computationally efficient, since the optimal control input (\rline{end} in \ralg{explorationalg}) can be obtained by evaluating the acquisitions only for the next states. 
A disadvantage of this approach, however, is that it might not guarantee an effective exploration for the \textit{whole} state space $X$, since the copter evaluates the acquisitions only locally. 
Another exploration strategy would be therefore to select the optimal state to be visited by evaluating the acquisitions for \textit{all} states in $X$ (instead of only locally for the next states), and then collect the sensor measurements to update the corresponding environmental beliefs. This leads us to a \textit{global selection-based approach}, and the details are given below.

\smallskip
\textcolor{black}{
\textit{(Global selection-based policy):}}  
{In the global selection-based approach, the copter first selects the optimal state that provides the highest acquisition in the whole state-space $X$, i.e., 
\begin{align}\label{selectmaxstate}
x^* = \arg\max_{x' \in X}\ W (x'). 
\end{align}
Then, the copter computes the optimal policy $\mu^* _{c_2} : X \rightarrow U$ such that the probability of reaching $x^*$ is maximized, i.e., 
\begin{align}\label{computeoptimalpolicy}
\mu^* _{c_2} \in \arg\max_{\mu_c}\ {\rm Pr} [{\bf x}_{c, \mu_{c}}  \models \Diamond x^*],
\end{align}
where ${\bf x}_{c, \mu_{c}} \in X^\omega$ denotes the state trajectory of the copter by applying the policy $\mu _c$, and $\Diamond x^*$ indicates the property that the state trajectory reaches $x^*$ in finite time (which corresponds to the "eventually" operator), i.e., ${\bf x}_{c, \mu_{c}} = {x}^0 _{c, \mu_{c}}{x}^1 _{c, \mu_{c}}{x}^2 _{c, \mu_{c}}\cdots \models \Diamond x^*$ iff there exists $k \in \mathbb{N}$ such that ${x}^k _{c, \mu_{c}} = x^*$. \req{computeoptimalpolicy} can be indeed solve via value iteration algorithm; for details, see \rsec{valueiterationsec}. 
Then, the copter moves to $x^*$ from the current state by applying $\mu^* _{c_2}$ so as to collect the corresponding sensor measurements and update the environmental beliefs of the atomic propositions. Once $x^*$ is reached, the copter re-computes the new $x^*$ based on the updated environmental beliefs, and iterate the same procedure as above for the time period $T_c$. The overall exploration algorithm based on the global selection-based policy is summarized in \ralg{explorationalgselection}. As shown in the algorithm, the copter first finds the optimal state $x^*$ that provides the highest acquisition among the whole state space $X$, and compute the optimal policy to reach $x^*$ (\rline{startselection}--\rline{startselectiontwo}). The copter applies the optimal policy until it reaches $x^*$ and collects the corresponding sensor measurements (\rline{whilestart}--\rline{whileend}). Note that the copter collects not only the sensor measurements for $x^*$, but also the ones for the states that are traversed while reaching to $x^*$, aiming to enhance the efficiency of exploration. The above procedure is iterated for the time period $T_c$. The variable $n_{succ}$ in \ralg{explorationalg} counts the number of times when the copter successfully reaches $x^*$. }

\textcolor{black}{The global selection-based approach should require a heavier computation than the local selection-based approach, since it needs to find the optimal state by evaluating the acquisitions for the whole state-space $X$, as well as to compute the optimal policy to reach $x^*$ via a value iteration. However, the advantage of employing this approach is that we can guarantee the \textit{coverage} of exploration, i.e., by repeating \ralg{explorationalg}, the environmental belief of the atomic propositions converge to the appropriate values; for details, see \rsec{convergencesec}.} 



\begin{algorithm}[t]
{\textcolor{black}{
\textbf{Input} and \textbf{Output} are the same as \ralg{explorationalg}; \\ 
$\ell \leftarrow 0$, $n_{succ} \leftarrow 0$; \\  
\While {$\ell < T_c -1$}{
Compute $H$ and $W$ according to \req{entropy} and \req{acquisition}, respectively; \label{startselection} \\
Compute $x^* \in X$ and $\mu^* _{c2} : X \rightarrow U_c$ according to \req{selectmaxstate} and \req{computeoptimalpolicy}, respectively; \label{startselectiontwo}\\ 
\While {$x \neq x^*$ and $\ell < T_c-1$ } { \label{whilestart}
Apply $u^* _c = \mu^* _{c2} (x)$ and sample the next state $x_{next} \sim p_c (\cdot |x, u^* _c)$;\\ 
$x \leftarrow x_{next}$, \ $\ell \leftarrow \ell+1$; \\ 
\For{each $(x', ap) \in X \times AP_c$ with $\|x - x'\| \leq R^c _{ap}$ }{\label{start1}
Provide the corresponding observation: $Z^c _x (x' \models ap) = z$; \\
Update $\mathcal{B}(x' \models ap)$ {according to \req{bayes}.} 
}
}\label{whileend}
\If {$x = x^*$} {
$n_{succ} \leftarrow n_{succ} + 1$; 
}
}
\Return $x$, $\mathcal{B}$; }
    \caption{\textcolor{black}{$\mathit{Exploration}(x, \mathcal{B}, b_{\max}, T_c)$ (global selection-based exploration)}} \label{explorationalgselection}
    }
\end{algorithm}

\subsection{Mission execution phase} \label{missionexecutesec}
In this subsection, we propose a detailed algorithm of the rover's mission execution phase. 
\subsubsection{Product belief MDP}
Given the environmental beliefs of the atomic propositions in \req{beliefatomic}, let $\mathcal{B} (x \models \sigma)$ for $x \in X$, $\sigma \in 2^{AP}$ be the \textit{joint belief} that  all atomic propositions in $\sigma$ are satisfied in $x$, i.e., $\mathcal{B} (x \models \sigma) = \prod_{ap \in \sigma} \mathcal{B} (x \models ap)$. 
Moreover, let $\mathcal{B} (x \models \sigma \wedge \neg (AP\backslash \sigma))$ be the joint belief that all atomic propositions in $\sigma$ are satisfied in $x$ and all atomic propositions in $AP$ other than $\sigma$ (i.e., $AP \backslash \sigma$) are not satisfied in $x$, i.e.,
\begin{align}\label{balpha}
\mathcal{B} (x \models \sigma \wedge \neg (AP\backslash \sigma)) = \prod_{ap \in \sigma} \mathcal{B} (x \models ap) \prod_{ap \in AP\backslash \sigma} (1- \mathcal{B} (x \models ap)). 
\end{align}
For simplicity of presentation, let 
$\mathcal{B}_{alph} (x \models \sigma) = \mathcal{B} (x \models \sigma \wedge \neg (AP\backslash \sigma))$.
Moreover, let $\mathcal{A}_\phi = (Q, 2^{AP}, \delta, q_0, Q_f)$ be an FSA corresponding to $\phi$, and, for each $(q, q') \in Q \times Q$, denote by $en (q, q') \subseteq 2^{AP}$ a subset of input alphabets, for which the transition from $q$ to $q'$ is allowed: $en (q, q') = \{\sigma \in 2^{AP}\ |\ q' \in \delta (q, \sigma)\}$. 
In addition, given $x \in X$ and $q, q' \in Q$, we let 
\begin{align}\label{ben}
\mathcal{B}_{en} (x \models en (q, q')) = \sum_{\sigma \in en(q, q') }\mathcal{B}_{alph} (x \models \sigma). 
\end{align}
That is, $\mathcal{B}_{en} (x \models en (q, q'))$ represents the belief that $q$ makes the transition to $q'$ from the atomic propositions that are satisfied in $x$. Note that we have $\sum_{q' \in Post (q)} \mathcal{B}_{en} (x \models en (q, q')) = 1$, since the collection of all events (the set of atomic propositions) corresponding to all outgoing transitions from $q$ are all possible events that can occur, i.e., $2^{AP}$. For this clarification, see an example below. \\ 

\begin{figure}[t]
        \centering
        \includegraphics[width=5cm]{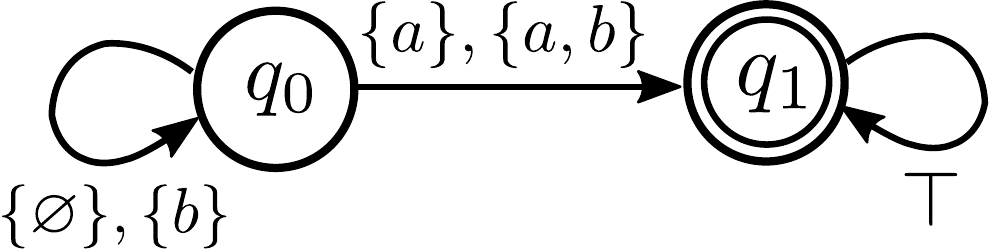}
        \caption{FSA that accepts all good prefix for the scLTL formula $\phi = \Diamond a$. The marked node represents the accepting state.}\label{automatonfig}
\end{figure}

\noindent 
\textit{(Example):}  
Consider the environment with two states $X= \{x_1, x_2\}$ and let $AP = \{ a, b \}$ and assume that the environmental beliefs of the atomic propositions are given by 
\begin{align}
&\mathcal{B} (x_1 \models a) = 0.1,\ \mathcal{B} (x_1 \models b) = 0.1,\ \mathcal{B} (x_2 \models a) = 0.9,\ \mathcal{B} (x_2\models b) = 0.2. 
\end{align}
Moreover, the scLTL formula is assumed to be given by $\phi = \Diamond a$. 
The corresponding FSA $\mathcal{A}_{\phi}$ that accepts all good prefix for $\phi$ is shown in \rfig{automatonfig}. 
For example, since $q_0 = \delta(q_0, \varnothing)$ and $q_0 = \delta(q_0, \{b\})$, we have $en (q_0, q_0) = \{\varnothing, \{b\}\}$. 
Moreover, we have 
\begin{align}
\mathcal{B}_{en} (x_1 \models en (q_0, q_0)) &= \mathcal{B}_{alph}(x_1 \models \varnothing) + \mathcal{B}_{alph}(x_1\models \{b\}) =0.9\times 0.9 + 0.9\times 0.1  = 0.9, 
\end{align}
which implies that, if the position of the rover is $x_1$, we have a high belief that $q_0$ provides the self-loop, i.e., the belief of reaching the accepting state $q_1$ is low. This is due to the fact that we have a low belief that $a$ is satisfied in $x_1$. 
Moreover, we have $\mathcal{B}_{en} (x_1, en (q_0, q_1)) = \mathcal{B}_{alph}(x_1, \{a\}) + \mathcal{B}_{alph}(x_1, \{a,b\}) = 0.09 + 0.01 = 0.1$. Note that we have $\mathcal{B}_{en} (x_1\models en (q_0, q_1)) + \mathcal{B}_{en} (x_1 \models en (q_0, q_0)) = 1$, 
satisfying the probabilistic nature. This is due to the fact that all outgoing transitions from $q_0$ are only $q_0$ and $q_1$, and the collection of all the corresponding events (set of atomic propositions) is $\{\varnothing, \{a\}, \{b\}, \{a, b\}\}$ ($=2^{AP}$), which is indeed all events that can occur. 
We also have $\mathcal{B}_{en} (x_2\models en (q_0, q_1)) = \mathcal{B}_{alph}(x_2\models \{a\}) + \mathcal{B}_{alph}(x_2\models \{a,b\}) = 0.72+0.18 =0.9$, implying that we have a high belief of reaching the accepting state, which is due to the fact that we have the high belief that $a$ is satisfied in $x_2$. \qedwhite 

\smallskip
Based on the above, we define the product belief MDP as a composition of the rover's motion model $\mathcal{M}_r$ and the FSA $\mathcal{A}_\phi$ as follows: 
\begin{mydef}\label{productmdp}
\normalfont 
Let $\mathcal{M}_r = (X, x_{r}, U_r, p_r)$ and $\mathcal{A}_\phi = (Q, 2^{AP}, \delta, q_0, Q_f)$ be the MDP motion model of the rover and the FSA corresponding to $\phi$, respectively. 
Moreover, given the environmental beliefs of the atomic propositions \req{beliefatomic}, let $\mathcal{B}_{en} (x \models en (q, q'))$ for $x \in X$, $q, q' \in Q$ be given by \req{ben}. 
Then, the product belief MDP between $\mathcal{M}_r$ and $\mathcal{A}_\phi$ is defined as a tuple $\mathcal{M}_S = (S, s_{0}, U_S, p_S, S_{f})$, where 
\begin{itemize}
\item $S = X \times Q$ is the set of states; 
\item $s_{ 0} =(x_{r}, q_0) \in S$ is the initial state;
\item $U_S = U_r$ is the set of control inputs; 
\item $p_S :S \times U_S \rightarrow \mathcal{D}(S)$ is the \textit{transition belief} function, defined as 
\begin{align}\label{transitionprod}
p _S ((x', q') | (x, q), u) = p_r (x' | x, u) \mathcal{B}_{en} (x \models en (q, q')); 
\end{align}
\item $S_{f} \subseteq S$ is the set of accepting states, where $S_{f} = X \times Q_f$. 
\qedwhite 
\end{itemize}
\end{mydef}
As shown in \req{transitionprod}, the transition function is called a \textit{belief} function instead of a probability function, which is due to that it is computed based on the environmental beliefs of the atomic propositions (i.e., the posterior given the past observations). 
As previously mentioned, $\mathcal{B}_{en} (x \models en (q, q'))$ represents the belief that $q$ makes the transition to $q'$ according to the atomic propositions that are satisfied in $x$. Hence, $p_S ( (x', q') | (x, q), u)$ indicates the joint belief that the pair $(x, q)$ makes the transition to $(x', q')$ by applying $u$.

\textcolor{black}{As shown in \rdef{productmdp}, the product MDP has the set of states involving only $X$ and $Q$. This leads to the reduction of the time complexity of synthesizing the optimal policy for the rover in contrast to the previous work; for details, see \rsec{valueiterationsec} (in particular \rrem{compurationalcomplexity}). } 


\subsubsection{Value iteration}\label{valueiterationsec}
Given $\mathcal{M}_S$, we denote by $\mu_S : S \rightarrow U_S (=U_r)$ a policy for $\mathcal{M}_S$, which associates a control input from $U_S$ for each state in $S$. Then, let $\mu_{S,seq}=\mu_S \mu_S \mu_S \ldots$ be the corresponding stationary policy sequence. 
We denote by ${\bf s}_{\mu_{S,seq}} = s(0) s(1) \ldots  \in S^\omega$ with $s(\ell) = (x(\ell), q(\ell))$, $\forall \ell \in \mathbb{N}_{\geq 0}$, the trajectory of the product belief MDP $\mathcal{M}_S$, such that $s(0) = s_0$ (i.e., $x(0) = x_r$, $q(0) = q_0$) and $s(\ell+1) \sim p_S (\cdot | s(\ell), \mu_S (s(\ell)))$ for all $\ell \in \mathbb{N}_{\geq 0}$. Given ${\bf s}_{\mu_{{S,seq}}} = s(0) s(1) \ldots$ with $s(\ell) = (x(\ell), q(\ell))$, $\forall \ell \in \mathbb{N}_{\geq 0}$, we can induce the corresponding trajectory of $\mathcal{A}_\phi$ as $q(0) q(1) \ldots \in Q^\omega$.  If the trajectory of $\mathcal{M}_S$ reaches the accepting state in $S_f$ in finite time, it means that the corresponding trajectory of $\mathcal{A}_\phi$ reaches the accepting state in $Q_f$ in finite time (i.e., it satisfies $\phi$). 
Hence, the problem of maximizing the belief for the satisfaction of $\phi$ defined by \req{beliefsatisfaction}, can be reduced to the problem of maximizing the belief that the trajectory of $\mathcal{M}_S$ reaches $S_f$ in finite time, i.e., 
\begin{align}\label{sprob}
\mu^* _S = \arg\max_{\mu_S}\ \mathcal{B}\left({\bf s}_{\mu_{S,seq}}  \models \Diamond S_f\right). 
\end{align}
The problem \req{sprob} can be indeed solved via a \textit{value iteration} as follows (see, e.g., \cite{abate2008,nilsson2018}). 
Let $\mathsf{1}_{S_{f}} : S \rightarrow \{0, 1\}$ be given by $\mathsf{1}_{S_{f}} (s) = 1$, if $s \in S_{f}$ and $0$ otherwise. Then, set $V^{0} (s) = \mathsf{1}_{S_{f}} (s)$, $\forall s \in S$ and for all $s \in S$, $\ell \in\mathbb{N}_{>0}$, 
\begin{align}
&V^{\ell+1} (s) = \max_{u \in U_S} \max \left(\mathsf{1}_{S_{f}} (s), \mathbb{E}_{s' \sim p_S (\cdot | s, u)} [V^{\ell} (s') | s, u] \right ), \label{valueiteone}\\ 
&\mu^{\ell+1} _S (s) = \arg\max_{u \in U_S} \max \left(\mathsf{1}_{S_{f}} (s), \mathbb{E}_{s' \sim p _S (\cdot | s, u)} [V^{\ell} (s') | s, u] \right ).  \label{valueitetwo}
\end{align} 
The above computations are given until they reach some fixed point, i.e., $\mu^{*} _S = \mu^{\ell'} _S = \mu^{\ell'+1} _S$ for some $\ell'$. Alternatively, one may iterate the above only for a given finite time steps $\overline{T} \in \mathbb{N}_{>0}$ with $\overline{T} \geq T_r$, i.e., iterate \req{valueiteone} and  \req{valueitetwo} for all $\ell \in \mathbb{N}_{0:\overline{T}-1}$ and set $\mu^{*} _S = \mu^{\overline{T}} _S$. 
This in turn implies to obtain the optimal policy that maximizes the belief that the trajectory of $\mathcal{M}_S$ reaches $S_f$ \textit{within the time interval} $[0, \overline{T}]$. 
Hence, it implies that we maximize the belief that the length of the good prefix of the word satisfying $\phi$ is less than $\overline{T}$. Given the optimal policy $\mu^* _S$ computed as above, we can induce the policy sequence for the rover based on the trajectory of $\mathcal{M}_S$, i.e., 
\begin{align}\label{roverpolicyseq}
\mu^* _{r, seq} = \mu^*_{r,0} \mu^*_{r,1} \mu^*_{r,2} \ldots, 
\end{align}
where $\mu^*_{r, \ell} (x(\ell)) = \mu^*_S (s(\ell))$, $\forall \ell \in \mathbb{N}_{\geq 0}$. 

\begin{myrem}\label{compurationalcomplexity}
\normalfont
\textcolor{black}{As shown in \rdef{productmdp}, the product MDP involves only $X$ and $Q$, and does not involve the states of the environmental beliefs as formulated in \cite{nilsson2018}. 
In particular, since the environmental beliefs of the atomic propositions are assigned for every state in $X$ in our problem setup, the time complexity of solving the value iteration algorithm is $O(|(E^{|X|} \times X \times Q)^2 \times U_S|)$ in \cite{nilsson2018} with $E$ being the set of states of the environment belief, while $O(|(X \times Q)^2 \times U_S|)$ in our approach. 
This implies that the time complexity of the value iteration algorithm in the previous work is exponential with respect to $|X|$, while it is polynomial in our approach. Therefore, our approach could alleviate the running time and the memory usage for synthesizing the optimal policies for the rover in contrast to the previous work.} \qedwhite  
\end{myrem}

\subsubsection{Computing the reachability belief and $b_{\max}$}\label{maxbelief}
Suppose that the current rover's position $x_r$ and the optimal policy $\mu^* _S$ is computed as above. Then, given $x \in X$  and $\ell \in \mathbb{N}_{0:T_r-1}$, we can compute a belief that the rover will reach $x$ from $x_r$ after $\ell$ time steps according to the optimal policy $\mu^* _S$. 
To this end, we denote the collection of all states of $\mathcal{M}_S$ by $S= \{s_1, s_2, s_3, \ldots, s_m\}$, where $m$ is the number of the states of $\mathcal{M}_S$. Given $x \in X$, we denote by $\mathcal{J}(x) \subseteq \{1, 2, \ldots, m\}$ the set of indices, for which the corresponding states of $\mathcal{M}_S$ include $x$, i.e., $\mathcal{J}(x) = \{ i \in \mathbb{N}_{1:m}\ |\ s_i = (x, q)\ {\rm for\ some}\ q\in Q\}$. 
If the policy $\mu^* _S$ is employed, the belief MDP $\mathcal{M}_S$ can be viewed as a \textit{belief Markov chain induced by} $\mu^* _S$, which is denoted by $\mathcal{M}^{\mu^* _S}_S = (S, s_0, p^{\mu^* _S} _S)$, where $S= \{s_1, s_2, s_3, \ldots, s_m\}$ is the set of states, $s_0 = (x_r, q_0)$ is the initial state, and $p^{\mu^* _S} _S$ is the transition belief function defined by $p^{\mu^* _S}_S (s' | s) = p_S (s' | s, \mu^* _S(s))$, $\forall s , s' \in S$. 
\begin{algorithm}[t]
{ 
\SetKwInOut{Input}{Input}
\SetKwInOut{Output}{Output}
\Input{$x_r$ (current rover's position); $\mathcal{B}$ (current environmental beliefs of the atomic propositions); $T_r$ (time period for the mission execution);} 
\Output{$x_r$ (updated current position); $b_{\max}$ (the mapping that represents maximum probability); $\mathcal{B}$ (updated environmental beliefs of the atomic propositions);}
Solve the value iteration \req{valueiteone}, \req{valueitetwo} until reaching some fixed point (or iterate them for all $\ell \in \mathbb{N}_{0:\overline{T}-1}$ for a given $\overline{T} \geq T_r$) and obtain the optimal policy $\mu^* _S : S \rightarrow U_S (=U_r)$; \\ 
$x \leftarrow x_r$,\ $q \leftarrow q_0$; \\ 
\For {$\ell = 0:T_r -1$}{
$u^* _r \leftarrow \mu^* _S (x, q)$ and sample $(x_{next}, q_{next}) \sim p_S (\cdot | (x, q), u^* _r)$; \\
$x \leftarrow x_{next}$, \ $q \leftarrow q_{next}$; \\
\For{each $(x', ap) \in X \times AP_r$ with $\|x - x'\| \leq R^r _{ap}$}{\label{belieupdatestart}
Provide the corresponding observation $Z^r _x (x', ap) = z$; \\
Update the belief as: $\mathcal{B}(x' \models ap) \longleftarrow \cfrac{{\rm Pr}[Z^r _x (x', ap) = z | x' \models ap]\mathcal{B}(x' \models ap)}{{\rm Pr}[Z^r _x (x', ap) = z]}$;  
}\label{belieupdateend}
}
$x_r \leftarrow x$ and \textcolor{black}{solve the value iteration \req{valueiteone}, \req{valueitetwo} to update the optimal policy $\mu^* _S$;}  \\ 
Compute $b_{\max}$ according to \rsec{maxbelief}; \\ 

\Return $x_r$, $b_{\max}$, $\mathcal{B}$; 
    \caption{$\mathit{MissionExecution}(x_r, \mathcal{B}, T_r)$ (main algorithm for mission execution)} \label{executionalg}
    }
\end{algorithm}
Now, let $b_{\ell} \in [0, 1]^m$ for all $\ell \in\mathbb{N}_{0:L-1}$ be recursively given by $b_{\ell+1} = A b_{\ell}$, where $A \in [0, 1]^{m\times m}$ is the transition matrix for the belief Markov chain $\mathcal{M}^{\mu^* _S}_S$, and $b^{(i)}_{0} = 1$ if $s_i$ is the initial state (i.e., $s_i = s_0 = (x_r, q_0)$) and $b^{(i)}_{0} = 0$ if $s_i$ is not the initial state. 
That is, $b^{(i)}_\ell$ represents a belief that the state $s_i$ is reached after $\ell$ time steps from the initial state $s_0 = (x_r, q_0)$ according to the optimal policy $\mu^* _S$. 
Based on the above, for each $x \in X$, we can compute a belief that $x$ is reached after $\ell$ time steps, denoted by $b_{\ell} (x)$, as $b_{\ell} (x) = \sum_{i \in \mathcal{J}(x)} b^{(i)} _\ell$. 
Then, let $b_{\max} (x)$ be given by $b_{\max} (x) = \max_{\ell \in \mathbb{N}_{0:T_r -1}} \ b_{\ell} (x)$,
i.e., $b_{\max} (x)$ indicates the maximum belief that the rover will reach $x$ (starting from $x_r$) within the time steps $T_r$. That is, for a large value of $b_{\max}(x)$ ($b_{\max}(x) \approx 1$), we have a high belief that the rover will reach $x$ at some point in the time interval $[0, T_r]$. 

As previously described in \rsec{decisionmaking}, the mapping $b_{\max}$ is utilized for the copter's exploration, so as to effectively search cells that are relevant to the mission execution. 

\subsubsection{Overall mission execution algorithm}
We now summarize the main algorithm of the mission execution phase in \ralg{executionalg}. As shown in the algorithm, the rover computes the optimal policy by solving the value iteration and apply it for the time period $T_r$. Moreover, while applying this policy, it takes the sensor measurements and updates the environmental beliefs of the atomic propositions \req{beliefatomic} (\rline{belieupdatestart}--\rline{belieupdateend}). Afterwards, the rover computes the mapping $b_{\max}$ according to the procedure described in \rsec{maxbelief}. 
Finally, the algorithm returns the current rover's position $x_r$, the mapping $b_{\max}$ and the updated belief for the atomic propositions in \req{beliefatomic}.\\

\textcolor{black}{
\section{{Convergence analysis}}\label{convergencesec}}
In this section, we analyze convergence property of the proposed algorithm presented in the previous section. 
In particular, we show that, by executing \ralg{alg} with the exploration phase given by the global selection-based approach (\ralg{explorationalgselection}), the environmental beliefs of the atomic propositions converge to the appropriate values, i.e., for all $x \in X$ and $ap \in AP$, $\mathcal{B}(x \models ap) \rightarrow 1$ if $ap \in L(x)$, and $\mathcal{B}(x \models ap) \rightarrow 0$ if $ap \notin L(x)$. Suppose that \ralg{alg} is implemented with the exploration phase given by \ralg{explorationalgselection}. 
To simplify the analysis, we make the following assumptions: 
\begin{myas}\label{copterreachassumption}
\normalfont 
For every execution of \ralg{explorationalgselection}, it follows that $n_{succ} \geq 1$. \qedwhite 
\end{myas}
\begin{myas}\label{copterbeliefassumption}
\normalfont 
For the sensor model of the copter and the rover, we assume that: 
\begin{enumerate}
\renewcommand{\labelenumi}{(\roman{enumi})}
\item $AP_c = AP_r = AP$. 
\item $R^c _{ap} = R^r _{ap} = 0$ for all $ap \in AP$. 
\item $\beta^c _x(x, ap) = \beta^c _x(x, ap') = \beta^r _x(x, ap) = \beta^r _x(x, ap')$ for all $\{ap, ap'\} \in AP \times AP$. \qedwhite
\end{enumerate} 
\end{myas}
\ras{copterreachassumption} excludes the case where \ralg{explorationalgselection} is terminated without reaching any selected states $x^*$ to be explored. Moreover, the first assumption in \ras{copterbeliefassumption} means that both the copter and the rover are equipped with all sensors for $AP$. The second assumption in \ras{copterbeliefassumption} means from \req{beta1} and \req{beta2} that the copter and the rover can only take the sensor measurements only on their current states. 
The third assumption in \ras{copterbeliefassumption} means that the precision of the sensor is the same for both the rover and the copter. 

For simplicity, we let $\beta = \beta^c _x(x, ap) = \beta^c _x(x, ap') = \beta^r _x(x, ap) = \beta^r _x(x, ap')$ for all $\{ap, ap'\} \in AP \times AP$. Note that $\beta > 0.5$ (for this clarification, see \req{beta1} and \req{beta2}). 
In addition, let $N_x \in \mathbb{N}_{>0}$ denote the total number of times the copter/rover visits the state $x$ and takes the corresponding sensor measurements for each $ap \in AP$, and let $m^{ap} _{N_x} \leq N_x$ ($x \in X$, $ap \in AP$) denote the number of times the corresponding sensor measurements for $ap$ are $1$. 
 In other words, $N_x - m^{ap} _{N_x}$ represents the number of times the corresponding observations for $ap$ are $0$. 
Finally, we make the following assumption: 
\begin{myas}\label{sensorprecisionas} 
\normalfont 
There exist $\varepsilon >0$ with $2\beta - 2 \varepsilon -1 >0$ and $\bar{N} \in \mathbb{N}_{>0}$, such that for all $x \in X$, $ap \in AP$ and $N_x \geq \bar{N}$, we have $\beta - \varepsilon \leq {m^{ap} _{N_x}}/{N_x} \leq \beta + \varepsilon$ if $ap \in L(x)$ and $\beta - \varepsilon \leq ({N_x - m^{ap} _{N_x}})/{N_x} \leq \beta + \varepsilon$ if $ap \notin L(x)$. \qedwhite
\end{myas}
\ras{sensorprecisionas} implies that the ratio between the number of sensor measurements and the number of making the correct sensor measurements (i.e., ${m^{ap} _{N_x}}/{N_x}$ if $ap \in AP$ and $({N_x - m^{ap} _{N_x}})/{N_x}$ if $ap \notin AP$) is $\varepsilon$-close to $\beta$ if the number of the visits (the sensor measurements) at $x$ is sufficiently large. 
The following theorem shows that, by executing \ralg{alg} with the exploration phase given by \ralg{explorationalgselection}, all the environmental beliefs of the atomic propositions converge to the appropriate values.  
\begin{mythm}\label{convergenceresult}
\normalfont
Let \ras{copterreachassumption}--\ref{sensorprecisionas} hold. 
Let $\alpha = 0$ in \req{acquisition} and suppose that \ralg{alg} is executed with the exploration phase given by \ralg{explorationalgselection}. 
Then, for all $x \in X$ and $ap \in AP$, it follows that 
\begin{numcases}
{\mathcal{B}(x \models ap) \rightarrow} 
1, \ \ {\rm if}\ ap \in L(x), \label{bconverge} \\ 
0, \ \ {\rm if}\ ap \notin L(x), \label{bconverge2}
\end{numcases}
as $k \rightarrow \infty$. \qedwhite
\end{mythm}
Recall that $k$ is defined in \ralg{alg} and represents the (global) time step during execution of \ralg{alg}. Hence, \rthm{convergenceresult} means that the environmental beliefs of the atomic propositions converge to the appropriate values as the number of the iterations for the exploration/mission execution phase goes to infinity. 
For the proof of \rthm{convergenceresult}, see Appendix \ref{proofoftheorem}.
\begin{myrem}
\normalfont
As shown in \rthm{convergenceresult}, the convergence properties \req{bconverge}, \req{bconverge2} may not hold if $\alpha \neq 0$ in \req{acquisition}. 
Nevertheless, as previously stated in \rsec{decisionmaking}, setting $\alpha \neq 0$ is useful and important for practical applications, since it can avoid the exploration of states that are completely irrelevant to the mission execution. Additionally, \req{bconverge}, \req{bconverge2} may not hold if the exploration phase is given by the local selection-based policy \ralg{explorationalg}. Nevertheless, as previously stated in \rsec{explorationsec}, utilizing \ralg{explorationalg} is useful for practical applications where the computation capacity of the copter is severely limited, since the optimal control input can be obtained by evaluating the acquisitions only for the next states. 
 \qedwhite 
\end{myrem}

Now, let $\hat{\mu}^* _{r, seq}$ denote the optimal policy sequence that maximizes the probability of satisfying $\phi$ under the assumption that the labeling function $L$ is \textit{known}, i.e., $\hat{\mu}^* _{r, seq} = \arg\min_{\mu_{r,seq}} {\rm Pr} [{\bf x}_{\mu_{r,seq}} \models \phi]$. If $L$ is known, $\hat{\mu}^* _{r, seq}$ can be derived by constructing a product MDP with the knowledge about $L$ and solving the value iteration algorithm (for details, see the proof of \rcol{corollary} in Appendix~B). Note that ${\mu}^* _{r, seq}$ given by \req{roverpolicyseq} is not necessarily equal to $\hat{\mu}^* _{r, seq}$, since ${\mu}^* _{r, seq}$ is derived by maximizing the \textit{belief} of satisfying $\phi$ based on the sensor measurements (i.e., ${\mu}^* _{r, seq} = \arg\min_{\mu_{r,seq}} \mathcal{B} ({\bf x}_{\mu_{r,seq}} \models \phi)$).  
The following Corollary is derived from \rthm{convergenceresult}, showing that ${\mu}^* _{r, seq}$ converges $\hat{\mu}^* _{r, seq}$ as $k \rightarrow \infty$. 
\begin{mycol}\label{corollary}
\normalfont
Let \ras{copterreachassumption}--\ref{sensorprecisionas} hold. 
Let $\alpha = 0$ in \req{acquisition} and suppose that \ralg{alg} is executed with the exploration phase given by \ralg{explorationalgselection}. Then, it follows that ${\mu}^* _{r, seq} \rightarrow \hat{\mu}^* _{r, seq}$ as $k \rightarrow \infty$. \qedwhite 
\end{mycol}
For the proof, see Appendix~B. 

\textcolor{black}{
\section{{Simulation results}}\label{simulationsec}}

In this section, we illustrate the effectiveness of the proposed algorithm through numerical simulations.
The simulation was conducted on \textcolor{black}{Python 3.7.9 with AMD Ryzen 7 3700U with Radeon Vega Mobile Gfx CPU and 16GB RAM}. 

\smallskip
\subsection{Simulation~1}

\subsubsection{Problem setup}
\begin{figure}
    \centering
    \includegraphics[width=5.8cm]{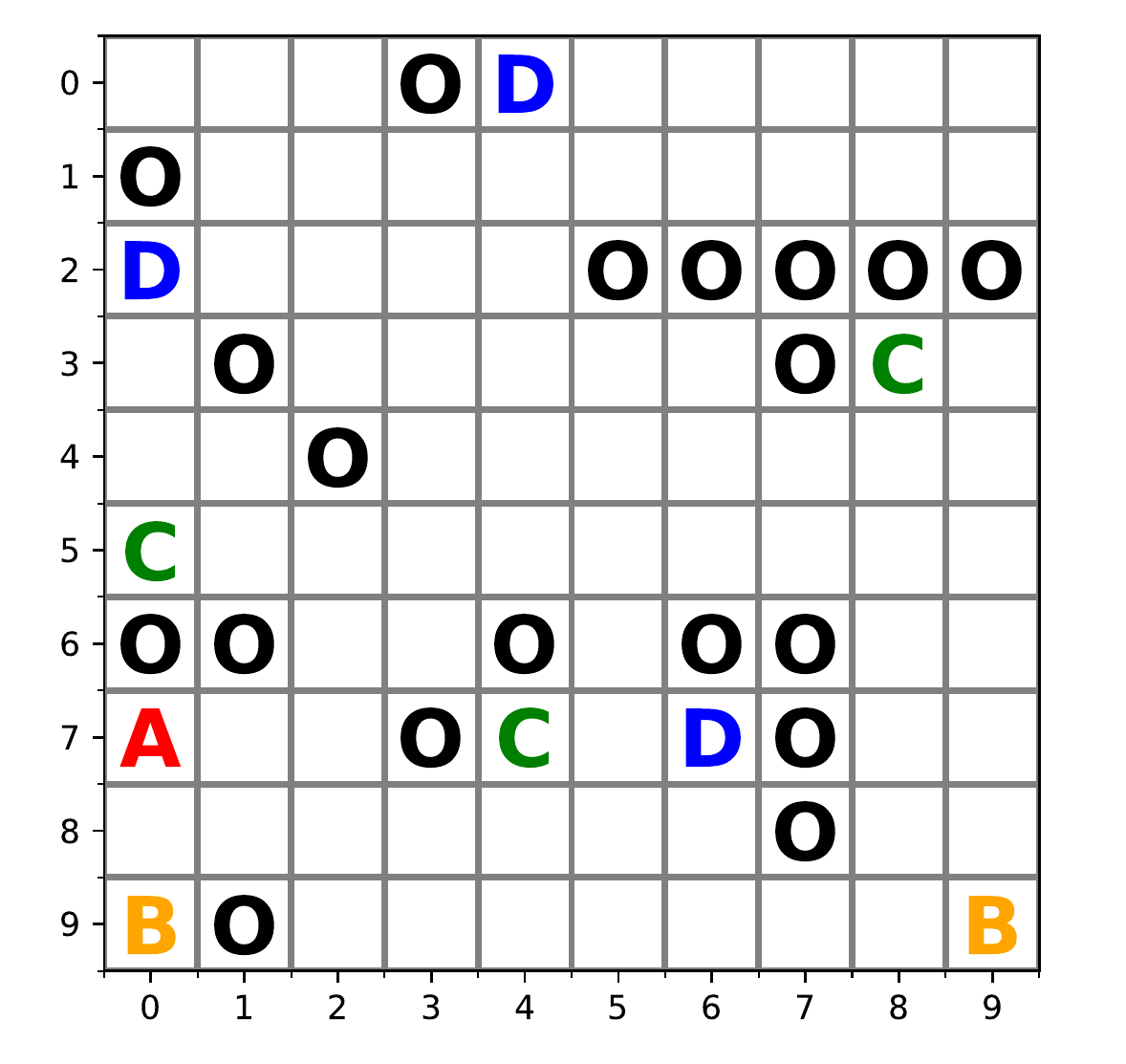}
    \caption{Environmental map of Simulation~1. Each alphabet (A,B,C,D and O) labeled in the cell indicates the atomic proposition that holds true in that cell.} 
    \label{sim1envmap}
\end{figure}
We consider the environmental map consisting of $n = 100$ cells as shown in \rfig{sim1envmap}. The set of states of the environment (i.e., the positions or the centroids of the cells) is given by \textcolor{black}{$X = \{[i, j]^\mathsf{T} \in \mathbb{R}^2, i = 0, \ldots, 9, j = 0, \ldots, 9 \}$}. The set of the atomic propositions is given by $AP = \{A, B, C, D, O\}$, where $A,B,C$ and $D$ represent the atomic propositions of \textit{target objects} that the rover seeks to discover, and $O$ represents the atomic proposition of \textit{obstacles} that the rover needs to avoid for all times. 
For both the MDP motion models of the rover and the copter, the set of states is given by $X$. The sets of control inputs for both the rover and the copter, i.e., $U_r$, $U_c$, are also the same and consists of $5$ components: \textit{stay in the same cell}, \textit{move up}, \textit{move down}, \textit{move right}, and \textit{move left}. Each control input drives the rover/copter towards one of the 8 cells adjacent to the current rover/copter's position. The transition probability for the rover's MDP is that there is $95$\% chance to move from the current cell to the desired cell, and there is the remaining 5\% chance to move to one of the cells adjacent to the desired cell (with equal probability). Regarding the copter's MDP, it is assumed that there is $90$\% chance to move from the current cell to the desired cell, and $10$\% chance to move to one of the cells adjacent to the desired cell. 
We assume that the copter is able to move without caring the obstacles (i.e., it can move on all states in $X$). 

The rover is equipped with sensors to detect both the target objects and the obstacles, i.e., $AP_r =\{A,B,C,D,O\}$, while the copter is equipped with a sensor to detect only the obstacles, i.e., $AP_c = \{O\}$. Moreover, the sensor range is assumed to be given by $R^r _{O} = R^r _{A} =R^r _{B}=R^r _{C}=R^r _{D}= 2 $, which implies from \req{beta1}, \req{beta2}, that the rover can provide high reliable sensor measurements only on its current cell and can provide row reliable measurements on its adjacent cells, and $R^c _{O} = 4$, which implies that the copter can detect obstacles within 4 cells far from the current cell. In addition, it is assumed that $M^r _{O}= M^r _{A} =M^r _{B}=M^r _{C}=M^r _{D}= 0.5$, which implies from \req{beta1}, \req{beta2} that the rover's maximum sensor accuracy is 100\%. Moreover, $M^c _{O}= 0.4$, which implies that the copter's maximum sensor accuracy is 90\%. The scLTL specification for the rover is given by 
\begin{align}\label{sim1ltl}
   \phi = \phi_1  \vee \phi_2 \vee \phi_3,
\end{align}
where 
\begin{align}
    \phi_1 = F_o A,\ 
    \phi_2 =  F_o B \wedge \bigcirc F_o C,\ \phi_3 = F_o C \wedge \bigcirc F_o D, 
\end{align}
with $F_o ap = \neg O {U} (\neg O \wedge ap)$ for $ap \in AP$. 
Intuitively, $\phi_1$ means that the rover should eventually discover the target $A$ while avoiding the obstacles. Moreover, $\phi_2$ (resp. $\phi_3$) means that the rover should eventually discover $B$ and then $C$ (resp. $C$ and then $D$) while avoiding the obstacles.  
During execution of \ralg{alg}, we set $T_c = 5$, $T_r = 3$, and $\alpha=1.5$ in the acquisition function of \req{acquisition}. Moreover, we assume that the mission is {(regarded as) complete} if the belief of satisfying $\phi$ by the rover's optimal policy exceeds $0.98$. 

\subsubsection{Simulation results}
\begin{figure}
    \centering
    \includegraphics[width=13cm]{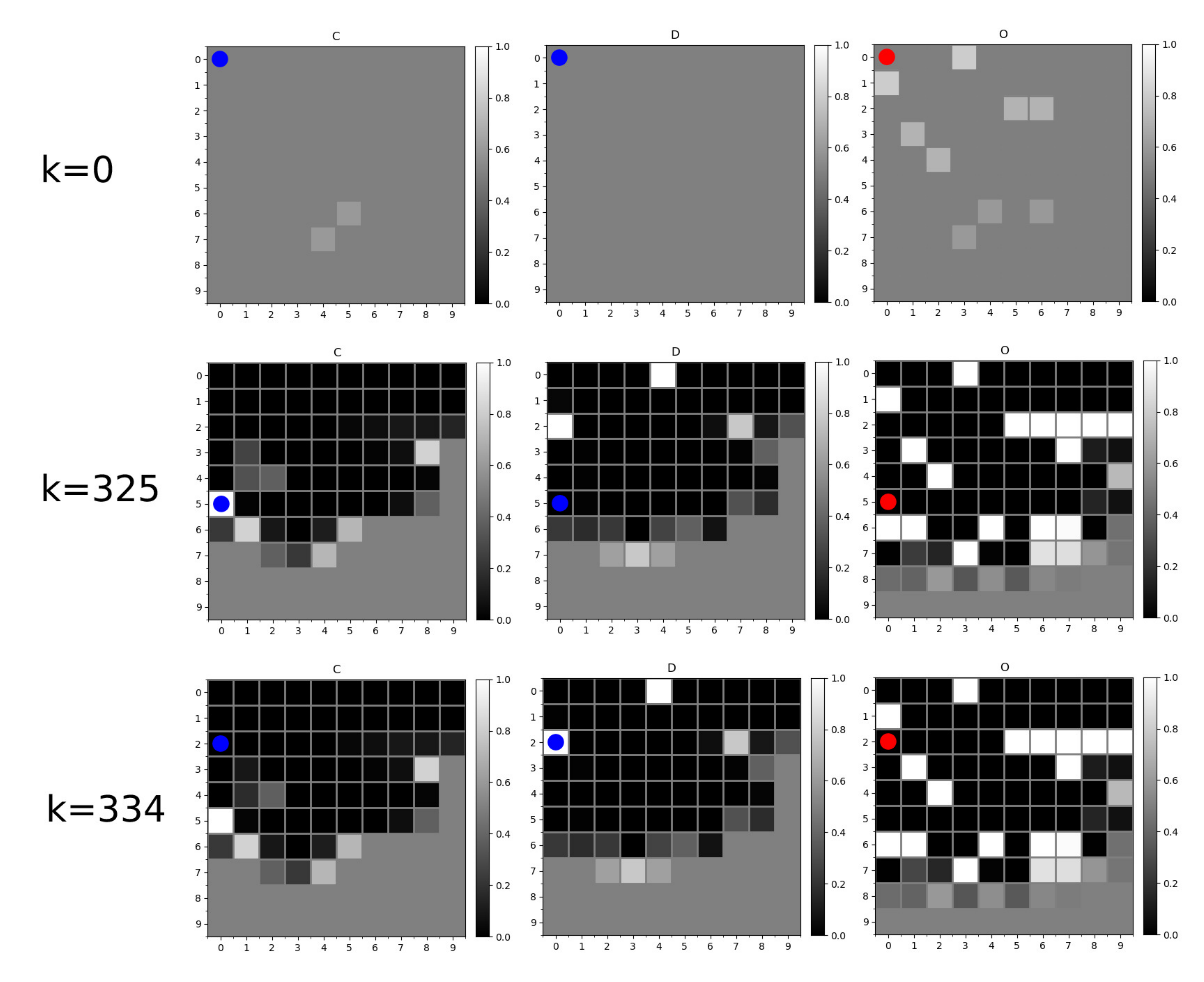}
    \caption{Snapshots of the environmental belief in Simulation~1. For simplicity, only the color maps for three propositions $C,D,O$ are shown. The blue circles represent the rover's position, and the red circles represent the copter's position. 
    The rover has reached the cell $(0,5)$ and found $C$ at $k=325$, and then reached $(0,2)$ where $D$ exists at $k=334$, regarding that the mission is complete. 
    It can be also verified that the rover has avoided obstacles for all times; for details, see the animation in \cite{animation}.} 
    \label{sim1snap}
\end{figure}
Some snapshots of the simulation result by applying \ralg{alg} are shown in \rfig{sim1snap}. During execution of \ralg{alg}, it is assumed that the copter executes the global selection-based policy (\ralg{explorationalgselection}) until the mission is complete. In the figure, the environmental beliefs of the atomic propositions are illustrated as the color maps, and, for simplicity, only the color maps for $C, D, O$ are shown. The rover's position is illustrated as the blue circle (only shown in the figures of $C, D$), and the copter's position is illustrated as the red circle (only shown in the figures of $O$). It can be shown from the figures that the rover has reached the cell where $C$ exists (at $k= 325$), and then reached the cell where $D$ exists (at $k=334$), regarding that the mission is complete. The whole behaviors of both the rover and the copter as well as the time elapse of the environmental beliefs of the atomic propositions can be shown in the animation; see \cite{animation}. 

To make comparisons between the local (\ralg{explorationalg}) and the global (\ralg{explorationalgselection}) selection-based exploration for the copter, we have iterated the following steps: (i) The initial positions of the rover and the copter are randomly chosen from $X$; (ii) Using the generated initial positions, execute \ralg{alg} with the local selection-based exploration (\ralg{explorationalg}); (iii) Using the generated initial positions, execute \ralg{alg} with the global selection-based exploration (\ralg{explorationalgselection}). The above steps have been iterated for 100 times, and for each exploration policy, we counted the number of times when the mission was successfully complete before $k=300$. At the same time, we also measured the average running time (in sec) of the local/global exploration policy (i.e., the average execution time of \ralg{explorationalg} and \ralg{explorationalgselection}). 
\rtab{resulttabone} illustrates the simulation results. 
The table indicates that the number of completing the mission via the global selection-based exploration is larger than the local one. This may be due to that the global selection-based exploration guarantees the convergence of the environmental beliefs of the atomic propositions, while the local one does not (see \rsec{overallexplorationsec}, \rsec{convergencesec}). 
On the other hand, the table also shows that the global one requires heavier computation than the local one, which is due to that it needs to solve the value iteration algorithm to reach the selected state with the highest acquisition (for details, see \req{computeoptimalpolicy}). 

\begin{table}[t]
\begin{center} 
\caption{The number of times when the mission was successfully complete for the local/global exploration policy, and the average running time (in sec) of executing the local/global exploration policy for each iteration in \ralg{alg}.} \label{resulttabone}
{\small 
\begin{tabular}{ccc} 
                          & Number of completing the mission & Average running time (s) \\ \hline \hline
Local  (\ralg{explorationalg}) & 62  & 3.0   \\ \hline
Global (\ralg{explorationalgselection}) & 71 & 31  \\ 
\hline \hline
\end{tabular}
}
\end{center}
\end{table}

\smallskip
\smallskip
\noindent

\color{black}
\subsection{Simulation 2: comparison with the existing algorithm}
In this section, we show that the proposed algorithm is advantageous over the existing algorithm \cite{nilsson2018} in terms of the running time and the memory usage of solving the value iteration (see \rrem{compurationalcomplexity} for the detailed explanation). We consider the environment with different size of the state space: $n = |X| \in \mathcal{N} = \{6, 9, 12, 15, 50, 100\}$. The set of the atomic propositions is given by $AP=\{A, O\}$, where $A$ indicates the target object and $O$ indicates the obstacle. The mission specification is given by $\phi = \neg O \ {U}\ (\neg O \wedge A)$. 
For each $n \in \mathcal{N}$, we randomly generate the initial beliefs of the atomic propositions $\mathcal{B}(x \models ap)$ for all $x \in X, ap \in AP$ uniformly from the interval $(0, 1)$ and solve the value iteration algorithm to synthesize the optimal policy for the rover according to the proposed approach (see \rsec{missionexecutesec}, in particular, \req{valueiteone} and \req{valueitetwo}).  
For the implementation of \cite{nilsson2018}, we assign the environmental belief for every state in the environment and solve the corresponding value iteration. 
The copter's exploration has not been given in this simulation, since we would like to focus on evaluating the running time of synthesizing the optimal policy for the rover. 
\rtab{resulttabsimtwo} shows the resulting running time (in sec) of solving the value iterations, where the symbol "---" indicates that the optimal policy could not be found due to the overflow of the memory. 
The table shows that the running time of solving the value iteration with the existing algorithm increases rapidly as $n$ increases and, in particular, it becomes infeasible when $n > 12$ due to the overflow of the memory \footnote{Note that in the numerical simulation in \cite{nilsson2018} considers the state space with $n=100$. This is due to that the atomic propositions are assigned \textit{only for some small regions in the state space}. Specifically, the numerical simulation in \cite{nilsson2018} considers that only $8$ or $5$ regions in the state-space are of interest to be explored (see Fig. 6 in \cite{nilsson2018}), so that the reduction of the computational complexity of solving the value iteration is achieved. However, as can be seen in our problem setup, we assume to assign the environmental beliefs for the atomic propositions \textit{for every single cell in the state space}. Thus, the algorithm in \cite{nilsson2018} has becomes infeasible for $n > 12$ in our problem setup.}. 
As described in \rrem{compurationalcomplexity}, such blowup is due to the fact that the size of the state-space of the product MDP increases exponentially with respect to $n$ $(= |X|)$. Therefore, the proposed approach is shown to be more useful than the existing approach in terms of the running time and the memory usage for synthesizing the optimal policy for the rover. 

\begin{table}[t]
\begin{center} 
\caption{Running time of solving the value iterations (in sec) using the proposed approach and the existing algorithm in \cite{nilsson2018}.} \label{resulttabsimtwo}
{\small 
\begin{tabular}{ccccccc} 
    $n$                  & 6 & 9 & 12 & 15 & 50 & 100\\ \hline \hline
Proposed approach                  & 0.02 & 0.05 & 0.13 & 0.22 & 3.91 & 20.35 \\ \hline
Previous approach in \cite{nilsson2018} & 0.02 & 1.07 & 86.02 & --- & --- & ---\\ 
\hline \hline
\end{tabular}
}
\end{center}
\end{table}

\section{Conclusion and future works}\label{conclusionsec}
In this paper, we investigate a collaborative rover-copter path planning and exploration with temporal logic specifications under uncertain environments. Mainly, the rover has the role to satisfy a mission specification expressed by an scLTL formula, while the copter has the role to assist the rover by exploring the uncertain environment and reduce its uncertainties. The environmental uncertainties are captured by the environmental beliefs of the atomic propositions, which represent the posterior probabilities that evaluate the level of uncertainties based on the sensor measurements. A control policy of the rover is then synthesized by maximizing a belief for the satisfaction of the scLTL formula through the implementation of an automata-based model checking. Then, an exploration policy for the copter is synthesized by evaluating the entropy that represents the level of uncertainties and the rover's path according to the current optimal policy. Finally, the effectiveness of the proposed approach is validated through several numerical examples. \textcolor{black}{Future works involve investigating safety guarantees (i.e., the rover avoids obstacles for all times during execution of Algorithm~1), as well as utilizing more sophisticated sensor models than the simple Bernoulli-type sensor models considered in this paper. In addition, extending the proposed approach to a more real-time and concurrency-related techniques, such as those that synthesize a supervisor that determines the activity of both the rover and the copter, should be studied for our further investigations. }

\section*{Acknowledgement} 
This work was supported by JST ERATO Grant Number JPMJER1603, JST CREST Grant Number JPMJCR2012, Japan and JSPS Grant-in-Aid for Young Scientists Grant Number JP21K14184.  \\


\appendix 
\section{Proof of \rthm{convergenceresult}}\label{proofoftheorem}
Let us first rewrite the Bayesian update \req{bayes} by 
\begin{align}
\mathcal{B}_{j+1} (x \models ap)= \frac{{\rm Pr}[Z_x (x\models ap) = z | x \models ap]\mathcal{B}_{j} (x \models ap)}{{\rm Pr}[Z_x (x \models ap) = z]}
\end{align}
for $j \in \mathbb{N}_{\geq 0}$, where we let $Z_x (x\models ap) = Z^r _x (x\models ap)$ (resp. $Z_x (x\models ap) = Z^c _x (x\models ap)$) if the rover (resp. copter) provides the sensor measurement for $ap$ at $x$. That is, $\mathcal{B}_j (x \models ap)$, $j \in \mathbb{N}_{\geq 0}$ represents the environmental belief of $ap$ at $x$ computed after $j$ times visit (by either the rover or the copter) of $x$. 
If $Z_x (x \models ap) = 1$ at the $j+1$-th visit of $x$, the Bayesian update is given by
\begin{align}
&\mathcal{B}_{j+1} (x \models ap) = \cfrac{ \beta \mathcal{B}_{j} (x \models ap)}{\beta \mathcal{B}_{j} (x \models ap) + ( 1- \beta) (1- \mathcal{B}_{j} (x \models ap))}.
\end{align}
After some simple calculations, we then obtain 
$\tilde{\mathcal{B}}_{j+1}(x \models ap) = \frac{1-\beta}{\beta} \tilde{\mathcal{B}}_{j}(x \models ap)$, where we let $\tilde{\mathcal{B}}_{j}(x \models ap) = \frac{1}{\mathcal{B}_{j}(x \models ap)}-1$. On the other hand, if $Z_x (x \models ap) = 0$ at the $j+1$-th visit, it follows that $\tilde{\mathcal{B}}_{j+1}(x \models ap) =\frac{\beta}{1-\beta} \tilde{\mathcal{B}}_{j}(x \models ap)$. 
Therefore, we obtain 
\begin{numcases}
{\tilde{\mathcal{B}}_{j+1}(x \models ap) =} 
 \frac{1-\beta}{\beta} \tilde{\mathcal{B}}_{j}(x \models ap), \ \ {\rm if}\ Z_x (x \models ap) = 1, \label{beliefrecursive} \\ 
\frac{\beta}{1-\beta} \tilde{\mathcal{B}}_{j}(x \models ap), \ \ {\rm if}\ Z_x (x \models ap) = 0. \label{beliefrecursive2} 
\end{numcases}

Suppose that the rover/copter visits $x$ the total $N_x$ times and that $ap \in L(x)$. 
From \req{beliefrecursive}, \req{beliefrecursive2}, we have 
\begin{align*}
\tilde{\mathcal{B}}_{N_x}(x \models ap)  &= \left( \cfrac{1-\beta}{\beta} \right )^{2 m^{ap} _{N_x} - N_x}\tilde{\mathcal{B}}_{0}(x \models ap). 
\end{align*}
Note that $\tilde{\mathcal{B}}_{0}(x \models ap) = \frac{1}{\mathcal{B}_{0}(x \models ap)}-1 \in (0, \infty)$, since the initial belief of the atomic proposition is selected as $\mathcal{B}_{0}(x \models ap) \in (0, 1)$ (see line~2 in \ralg{alg}). 
From \ras{sensorprecisionas}, it follows that 
$N_x (2\beta - 2\varepsilon-1) \leq 2 m^{ap} _{N_x} - N_x \leq N_x (2\beta + 2\varepsilon-1)$ for all $N \geq \bar{N}$ and $ap \in L(x)$. 
Thus, we obtain $0< \tilde{\mathcal{B}}_{N_x}(x \models ap) \leq \{ ({1-\beta})/{\beta} \}^{N_x (2 \beta - 2 \varepsilon -1 )}\tilde{\mathcal{B}}_{0}(x \models ap) $
for all $N_x \geq \bar{N}$ and $ap \in L(x)$. 
Noting that $2 \beta- 2 \varepsilon -1  > 0$ (see \ras{sensorprecisionas}), $\beta > 0.5$ and $\tilde{\mathcal{B}}_{0}(x \models ap) \in (0, \infty)$, we obtain $\tilde{\mathcal{B}}_{N_x}(x \models ap) \rightarrow 0$ as $N_x \rightarrow \infty$, which implies that $\mathcal{B}_{N_x}(x \models ap) \rightarrow 1$ as $N_x \rightarrow \infty$. 
Similarly, we obtain $\tilde{\mathcal{B}}_{N_x}(x \models ap) \rightarrow \infty$, i.e., $\mathcal{B}_{N_x}(x \models ap) \rightarrow 0$ if $ap \notin L(x)$ as $N_x \rightarrow \infty$. 
Hence, for all $x \in X$ and $ap \in AP$, we have 
\begin{numcases}
{\lim _{N_x \rightarrow \infty} \mathcal{B}_{N_x} (x \models ap) =} 
1, \ \ {\rm if}\ ap \in L(x), \label{bconvergenx} \\ 
0, \ \ {\rm if}\ ap \notin L(x). \label{bconvergenx2}
\end{numcases}
In other words, \req{bconverge}, \req{bconverge2} are satisfied for all $x \in X$ and $ap \in AP$, if for all $x\in X$, the number of visits at $x$ goes to infinity as $k\rightarrow \infty$, i.e., $N_x \rightarrow \infty$ as $k\rightarrow \infty$. 
In what follows, it is shown that $N_x \rightarrow \infty$ as $k\rightarrow \infty$ for all $x\in X$. 
With a slight abuse of notation, let $N_{x} (k) \leq k$ and $m^{ap} _{N_x} (k) \leq N_{x} (k)$ denote, respectively, the number of total times the rover/copter visits $x$ within the time step $k$ and the number of times the corresponding observations for $ap$ are $1$. To show by contradiction, let $X_{not} \subset X$ denote the set of all states at which the number of visits does \textit{not} go to the infinity as $k \rightarrow \infty$, and assume that $X_{not}$ is non-empty. 
In other words, there exists a time step $k_{ter} \in \mathbb{N}_{>0}$ such that all $x \in X_{not}$ are no more visited after $k_{ter}$, i.e., $N_x (k) = N_x (k+1)$ for all $k \geq k_{ter}$. Hence, we have $\mathcal{B}_{N_x (k)} (x \models ap) = \mathcal{B}_{N_x (k+1)} (x \models ap) \in (0, 1)$ for all $k \geq k_{ter}$, and, therefore,  $H(\mathcal{B}_{N_x (k)} (x \models ap)) = H(\mathcal{B}_{N_x (k+1)} (x \models ap)) \in (0, 1)$ 
for all $k \geq k_{ter}$, which implies that the entropy remains constant and does \textit{not} converge to $0$. 
On the other hand, it follows that, for all $x' \in X \backslash X_{not}$, $N_{x'} (k) \rightarrow \infty$  as $k \rightarrow \infty$.
Hence, the environmental beliefs of the atomic proposition converge to the appropriate values, i.e., 
for all $x' \in X \backslash X_{not}$ and $ap \in AP$, $\mathcal{B}_{N_{x'} (k)} (x' \models ap) \rightarrow 1$ if $ap \in L(x')$ and $0$ if $ap \notin L(x')$ as $k\rightarrow \infty$. Therefore, the entropy converges to $0$, i.e., for all $ap \in AP$ and $x' \in X \backslash X_{not}$, $H(\mathcal{B}_{N_{x'} (k)} (x' \models ap)) \rightarrow 0$ as $k \rightarrow \infty$. 

Now, since $\sum_{ap \in AP} H(\mathcal{B}_{N_{x'} (k)} (x' \models ap)) \rightarrow 0$ for all $x' \in X \backslash X_{not}$, there exist a time step $\bar{k} \geq k_{ter}$ such that the following holds: for all $k \geq \bar{k}$, $x \in X_{not}$ and $x' \in X \backslash X_{not}$, 
\begin{align}
\sum_{ap \in AP} H(\mathcal{B}_{N_{x'} (k)} (x' \models ap)) < \sum_{ap \in AP} H(\mathcal{B}_{N_{x} (k)} (x \models ap)).  \label{compareentropy}
\end{align} 
Recalling that we set $\alpha = 0$ in \req{acquisition}, the inequality \req{compareentropy} implies that, at a certain time step after $\bar{k} \geq k_{ter}$, the copter would select some $x^* \in X_{not}$ at the beginning of execution of \ralg{explorationalgselection}, which means from \ras{copterreachassumption} that the copter {would have} visited $x^*$ after $\bar{k} \geq k_{ter}$. However, this contradicts the assumption that all $x \in X_{not}$ are not visited after $k_{ter}$. Overall, such contradiction follows from the fact that we assume $X_{not}$ is non-empty. Therefore, it follows that $X_{not}$ is empty, and thus $N_{x} (k) \rightarrow \infty$ as $k\rightarrow \infty$ for all $x \in X$. 
In summary, \req{bconverge} and \req{bconverge2} are satisfied for all $x \in X$ and $ap \in AP$. \qedwhite 

\section{Proof of \rcol{corollary}}
Let $\hat{\mathcal{M}}_S = ({S}, {s}_{0}, {U}_S, \hat{p}_S, {S}_{f})$ denote the product MDP between $\mathcal{M}_r = (X, x_{r}, U_r, p_r)$ and $\mathcal{A}_\phi = (Q, 2^{AP}, \delta, q_0, Q_f)$, where $S$, ${s}_{0}$, ${U}_S$ and $S_f$ are the same as the product belief MDP defined in \rdef{productmdp}, and $\hat{p}_S :S \times U_S \rightarrow \mathcal{D}(S)$ is the {transition probability} function, defined by 
$\hat{p}_S ((x', q') | (x, q), u) = p_r (x' | x, u)$ (for all $\{(x, q), (x', q')\} \in S \times S$ and $u \in U_S$ )
iff $L(x) \in en (q, q')$ and $0$ otherwise. 
If $L$ is \textit{known}, the optimal policy $\hat{\mu}^* _{r, seq}$ is then obtained by solving the value iteration algorithm over the product MDP $\hat{\mathcal{M}}_S$. This fact, combined with \rthm{convergenceresult}, implies that if the product \textit{belief} MDP ${\mathcal{M}}_S$ in \rdef{productmdp} converges to the product MDP $\hat{\mathcal{M}}_S$, i.e., ${\mathcal{M}}_S \rightarrow \hat{\mathcal{M}}_S$ as $k \rightarrow \infty$, then ${\mu}^* _{r, seq} \rightarrow \hat{\mu}^* _{r, seq}$ as $k \rightarrow \infty$. To show that ${\mathcal{M}}_S \rightarrow \hat{\mathcal{M}}_S$ as $k \rightarrow \infty$, we need to show that $p _S ((x', q') | (x, q), u) \rightarrow \hat{p}_S ((x', q') | (x, q), u)$ (for all $\{(x, q), (x', q')\} \in S \times S$ and $u \in U_S$). Suppose that $\mathcal{B}(x \models ap) \rightarrow 1$ (resp. $\mathcal{B}(x \models ap) \rightarrow 0$) for all $ap \in L(x)$ (resp. $ap \notin L(x)$), i.e., all the environmental beliefs of the atomic propositions converge to the appropriate values. 
From \req{balpha}, it then follows that $\mathcal{B}_{alph} (x \models \sigma) \rightarrow 1$ iff $\sigma = L(x)$ and $0$ otherwise. From \req{ben}, we then obtain $\mathcal{B}_{en} (x \models \sigma) \rightarrow 1$ iff $\sigma = L(x)$ and $0$ otherwise. Hence, the transition belief function \req{transitionprod} becomes $p _S ((x', q') | (x, q), u) \rightarrow p_r (x' | x, u)$ (for all $\{(x, q), (x', q')\} \in S \times S$ and $u \in U_S$ ) iff $L(x) \in en (q, q')$ and $0$ otherwise, which indeed coincides with $\hat{p}_S ((x', q') | (x, q), u)$. Hence, it follows that $p _S ((x', q') | (x, q), u) \rightarrow \hat{p}_S ((x', q') | (x, q), u)$ (for all $\{(x, q), (x', q')\} \in S \times S$ and $u \in U_S$ ), and, therefore, ${\mathcal{M}}_S \rightarrow \hat{\mathcal{M}}_S$ as $k \rightarrow \infty$. As described above, this directly follows ${\mu}^* _{r, seq} \rightarrow \hat{\mu}^* _{r, seq}$ as $k \rightarrow \infty$. \qedwhite
\end{document}